%% file: iot.tex
\documentclass[10pt,journal]{IEEEtran}

\usepackage{booktabs} 
\usepackage{graphicx}
\usepackage{adjustbox} 
\usepackage{mwe}
\graphicspath{{./img/} {./figs/} {./figs_primer/camera_ready/}}
\usepackage{pbox}
\usepackage{makecell}

\usepackage[hyphens]{url}
\usepackage{algorithm}
\usepackage{algpseudocode}

\usepackage{caption,subcaption}
\usepackage{mathtools}
\usepackage{multirow}
\usepackage{xspace}
\usepackage{etoolbox}
\usepackage{float}

\usepackage[shortcuts]{extdash}

\usepackage[colorinlistoftodos]{todonotes}
\usepackage{color,soul}

\usepackage[acronym]{glossaries}
\newacronym{3gpp}{3GPP}{Third Generation Partnership Project}

\newacronym{ue}{UE}{User Equipment}
\newacronym{nbiot}{NB-IoT}{Narrowband-IoT}
\newacronym{iot}{IoT}{Internet of Things}
\newacronym{mtc}{MTC}{Machine Type Communication}
\newacronym{lte}{LTE}{Long Term Evolution}
\newacronym{bs}{BS}{Base Station}
\newacronym{dl}{DL}{Downlink}
\newacronym{ul}{UL}{Uplink}
\newacronym{ack}{ACK}{Acknowledgment}
\newacronym{mcl}{MCL}{Maximum Coupling Loss}
\newacronym{ecl}{ECL}{Extended Coverage Level}
\newacronym{nrsrp}{NRSRP}{Narrowband Reference Signal Received Power}
\newacronym{rsrp}{RSRP}{Reference Signal Received Power}
\newacronym{rsrq}{RSRQ}{Reference Signal Received Quality}
\newacronym{snr}{SNR}{Signal to Noise Ratio}
\newacronym{sinr}{SINR}{Signal to Interference plus Noise Ratio}
\newacronym{rssi}{RSSI}{Received Signal Strength Indicator}
\newacronym{txpower}{TX Power}{Transmission Power}
\newacronym{rach}{RACH}{Random Access Procedure}
\newacronym{idrx}{iDRX}{Idle state DRX}

\newacronym{dci}{DCI}{Downlink Control Indicator}
\newacronym{kpis}{KPIs}{Key Performance Indicators}
\newacronym{ru}{RU}{Resource Unit}
\newacronym{sf}{SF}{Subframe}

\newacronym{tbs}{TBS}{Transmission Block Size}
\newacronym{rai}{RAI}{Release Assistance Indicator}

\newacronym{npusch}{NPUSCH}{Narrowband Physical Uplink Shared Channel}
\newacronym{rrc}{RRC}{Radio Resource Control}
\newacronym{phy}{PHY}{Physical}

\newacronym{drx}{DRX}{Discontinuous Reception}
\newacronym{psm}{PSM}{Power Saving Mode}
\newacronym{edrx}{eDRX}{Extended Discontinuous Reception}
\newacronym{cdrx}{cDRX}{Connected state Discontinuous Reception}
\newacronym{ptw}{PTW}{Paging Time Window}
\newacronym{tau}{TAU}{Tracking Area Update}

\newacronym{sr}{SR}{Service Request}
\newacronym{tx}{TX}{Transmission}
\newacronym{rx}{RX}{Reception}
\newacronym{rl}{RL}{Release}
\newacronym{pci}{PCI}{Physical Cell Identity}
\newacronym{rtt}{RTT}{Round Trip Time}

\newtoggle{anonSubmission}
\toggletrue{anonSubmission}

\iftoggle{anonSubmission}
{
    \newcommand{\telenor}{Op1\xspace}
    \newcommand{\telia}{Op2\xspace}
}
{
    \newcommand{\telenor}{Telenor\xspace}
    \newcommand{\telia}{Telia\xspace}
}

\iftoggle{anonSubmission}
{
      \newcommand{\ublox}{\textrm{SARA-N211}\xspace}
    \newcommand{\quectel}{\textrm{Quectel-BC95}\xspace}

}
{
    \newcommand{\ublox}{Modu_1\xspace}
    \newcommand{\quectel}{Modu_2\xspace}
  }

\newcommand{\eclZero}{\texttt{ECL: 0}\xspace}
\newcommand{\eclOne}{\texttt{ECL: 1}\xspace}
\newcommand{\eclTwo}{\texttt{ECL: 2}\xspace}

\title{Dissecting Energy Consumption of NB-IoT Devices Empirically}

\author{Foivos~Michelinakis,
Anas~Saeed~Al-selwi,
Martina~Capuzzo,
Andrea~Zanella,
Kashif~Mahmood,
Ahmed~Elmokashfi
\thanks{F. Michelinakis, A. Saeed Al-selwi and A. Elmokashfi are with
Simula Metropolitan, Oslo, 0167 Norway (e-mail: {foivos, anasal, ahmed}@simula.no).}
\thanks{M. Capuzzo and A. Zanella are with University of Padova
Italy (e-mail: {capuzzom, zanella}@dei.unipd.it).}
\thanks{K. Mahmood is with Telenor Research
Norway (e-mail: Kashif.Mahmood@telenor.com).}
}

\begin{document}

\newcommand{\TODO}[1]{\textcolor{red}{\textbf{TODO: #1}}}
\newcommand{\foivos}[1]{\textcolor{blue}{\textbf{foivos: #1}}}
\newcommand{\AEL}[1]{\textcolor{violet}{\textbf{AE: #1}}}
\newcommand{\AZ}[1]{\textcolor{green}{\textbf{AZ: #1}}}
\newcommand{\MC}[1]{{\textcolor{red}{MC: #1}}}

\newcommand\eg{\emph{e.g.},\xspace}
\newcommand\ie{\emph{i.e.},\xspace}
\newcommand\etc{\emph{etc}.\xspace}
\newcommand\via{\emph{via}}

\maketitle

\begin{abstract}

\input{abstract}

\end{abstract}



\input{intro}
\input{nbiot_primer}

\input{methodology}
\input{metadata_quality}
\input{connected}
\input{idle}
\input{rttPacketLoss}
\input{resultsSummary}

\input{discussion}
\input{related_work}
\input{conclusions}
\input{ack}

\bibliographystyle{IEEEtran}
\bibliography{iot}

\appendices
\input{dataPreprocessingAndMetadata}
\input{snr_rsrp_mapping}

\end{document}

%% file: abstract.tex

3GPP has recently introduced NB-IoT, a new mobile communication standard
offering a robust and energy efficient connectivity option to the rapidly
expanding market of Internet of Things (IoT) devices.
To unleash its full potential, end-devices are expected to work in a plug and
play fashion, with zero or minimal parameters configuration, still exhibiting
excellent energy efficiency.
We perform the most comprehensive set of empirical measurements with commercial
IoT devices and different operators to date, quantifying the impact of
several parameters to energy consumption.
Our campaign proves that parameters setting does impact energy consumption, so
proper configuration is necessary.
We shed light on this aspect by first illustrating how the nominal standard
operational modes map into real current consumption patterns of NB-IoT devices.
Further, we investigate which device reported metadata metrics better
reflect performance and implement an algorithm to automatically identify
device state in current time series logs.
Then, we provide a measurement-driven analysis of the energy consumption and
network performance of two popular NB-IoT boards under different parameter
configurations and with two major western European operators.
We observed that energy consumption is mostly affected by the paging interval in
Connected state, set by the base station.
However, not all operators correctly implement such settings.
Furthermore, under the default configuration, energy consumption in not strongly
affected by packet size nor by signal quality, unless it is extremely bad.
Our observations indicate that simple modifications to the default parameters
settings can yield great energy savings.

%% file: intro.tex
\section{Introduction}
\label{sec:intro}

The recent explosion in the number of IoT devices has been supported by a few
proprietary low power wide area systems, which rely on unlicensed spectrum.
Their popularity caused \gls{3gpp} to investigate cellular IoT technologies, resulting
in the development of \gls{lte}-M and \gls{nbiot} standards.
A main focus area of these technologies is high energy efficiency, enabling
devices operated by tiny batteries to operate for prolonged periods of time.
The main advantages of the \gls{3gpp} standards are the use of licensed spectrum and
the fact that they build upon existing \gls{3gpp} technologies, allowing for more
stable and predictable performance, and reuse of infrastructure.
\gls{lte}-M and \gls{nbiot} are critical in enabling future 5G networks to
support the density and latency requirements of massive machine type
communications~\cite{iot5g}.
They can also seamlessly coexist with the upcoming New Radio (NR) access
technology, since the latest standards allow the reservation of NR
time-frequency resources for \gls{lte}-M and \gls{nbiot} transmissions.
In this work, we focus on \gls{nbiot}, which provides lower throughput but more
robust connectivity than \gls{lte}-M, and is hence geared towards massive
deployments of IoT devices.


Energy efficiency is certainly a major concern for typical IoT deployment
scenarios, since batteries of IoT devices are not meant
to be recharged or replaced, tying the lifetime of the battery to the lifetime
of the device itself.
Our analysis aims to quantify the impact of several parameters to energy
consumption and reveals that network configurations may greatly affect the
device lifetime, without offering performance gains.
For example, we show that setting a flag may reduce the energy needed to transmit an 
Uplink packet and receive a response under good signal conditions from 0.82 J to
0.12 J, with no performance penalty.
In a scenario, where 6 messages per day are transmitted, the device's lifetime
is extended from 8.5 years to 30.
\gls{nbiot} users though, are naturally inclined to believe that, in
analogy with broadband cellular services, \gls{nbiot} services can also be
accessed in a plug--and--play fashion, without or with minimal set-up of the end
devices.
In the same fashion, application developers should not rely on default settings
and, instead, carefully pick parameter values that best match the tradeoff
between delay and device lifetime of their use-case.

The purpose of this paper is to go beyond the early studies of empirical
\gls{nbiot} performance
characterization, most notably ~\cite{martinez2019exploring,yeoh2018experimental,duhovnikov2019power},
whose findings and limitations are discussed in detail in Section~\ref{sec:discussion}.
Comparatively, our experiments are more comprehensive: we 1) test more operators
and / or more modules, thus revealing inefficiencies of specific module-operator
combinations 2) use the latest NB-IoT features and 3) study a bigger
variety of scenarios.
In particular, we analyze the intricacies of operator configuration and strategies that greatly
affect key metrics and battery life, while also deep diving into the performance
of Release-13 enhancements.
We conduct the first exhaustive experimental study of its kind,
exploring the \gls{nbiot} ecosystem, under various power management configurations.
The experiments involved two different \gls{nbiot} boards and two main
telecommunication operators in a western European country, so as to appreciate
the impact of implementation choices on the system energy efficiency.
Since we focus on parameter tuning the main findings can be generalized to
other networks and devices.
Our measurements are spread across several months in the period between October
2018 and October 2019.

Our main contributions are:
1) a thorough presentation of the power-saving
mechanisms supported by the latest commercially available \gls{nbiot} release;
2) experimental study of the different configurations and operator strategies,
where we quantify their impact on energy consumption and network \gls{kpis}
at the \gls{rrc} state level and, when applicable, within a state;
3) analysis of which metadata metrics better reflect the device behavior and
4) an algorithm for extracting the state of a device directly from
current time series logs.

In the sequel, we elaborate on some rather surprising findings.
The energy consumption of \gls{nbiot} devices does not seem
to be strongly affected by channel conditions, except in extremely harsh
conditions.
Furthermore, under default parameters setting, the packet size has negligible
impact on the overall device power consumption, while its effect becomes more
significant when energy saving mechanisms are used.
Based on such empirical observations, we provide indication for device-side and
network-side parameter configurations that yield similar application-level
performance, while preserving the device battery.
We have communicated the findings to the measured operators, and they
reconfigured their networks accordingly resulting in a boost in energy
efficiency for end users. 

The remainder of this paper is structured as follows:
Section~\ref{sec:nbiot_primer} is an introduction to the specifics and
mechanisms of \gls{nbiot} and Section~\ref{sec:methodology} describes our
experiment workflow.
Sections~\ref{sec:energy_consumption_connect}
and~\ref{sec:energy_consumption_idle} break down the parameters that affect
energy consumption, Section~\ref{sec:rttPacketLoss} discusses network \gls{kpis}
and Section~\ref{sec:resultsSummary}
summarizes our findings.
Section~\ref{sec:discussion} looks into how the above affect typical
\gls{nbiot} usecases,
Section~\ref{sec:related_work} condenses the
existing literature and Section~\ref{sec:conclusions} concludes this article.
Finally, we include two technical appendices, where we discuss how we isolate
device states and prove the relationship of two metadata metrics.

%% file: nbiot_primer.tex
\section{Primer on NB-IoT}
\label{sec:nbiot_primer}
\gls{nbiot} occupies a bandwidth of 180 kHz within the \gls{lte}
spectrum, according to three possible options: (i) Standalone, where \gls{nbiot}
is placed in existing idle spectrum resources, (ii) Guardband, where the
\gls{lte} guard bands are used for \gls{nbiot}, and (iii) In-band, where in-band
\gls{lte} resource blocks are assigned to \gls{nbiot}~\cite{liberg2019cellular}.
The main features of the technology are listed in Table~\ref{tab:nbiot-features}.
Next, we briefly present the main operational modes and power saving mechanisms
of \gls{nbiot}.


\begin{table}
  \centering
    \caption{Main features of NB-IoT~\cite{liberg2019cellular,sinha2017survey,
      mekki2019comparative}.}
   \label{tab:nbiot-features}
  \begin{tabular}{ll}
    \toprule
    \textbf{Feature} & \textbf{NB-IoT} \\
    \midrule
    Frequency & Licensed LTE frequency bands \\
    Bandwidth & 180~kHz \\
    \pbox{5cm}{Theoretical peak data rate\\
    at the physical layer}& 226.7 kbps (DL); 250 kbps (UL) \\
    Range & $\sim$1km (urban); $\sim$10km (rural) \\
    Handover & Not available\footnotemark  \\
    Latency & $\leq 10s$ \\
    Low power mechanisms & eDRX, PSM \\
    \bottomrule
    \vspace{-2em}
  \end{tabular}
\end{table}

\footnotetext{The handover functionality is not considered in the standard:
  therefore, a new connection procedure is required for mobile devices entering
  in a cell covered by a different \gls{bs}. Further, \gls{nbiot} does not
  officially support mobility. Cell reselection is intended only for attaching
  to a cell with better coverage.}

\subsection{Operational phases}
\label{sec:operational_phases}

\begin{figure*}[t]
  \centering
  \begin{subfigure}[t]{\textwidth}
    \includegraphics[width=\textwidth]{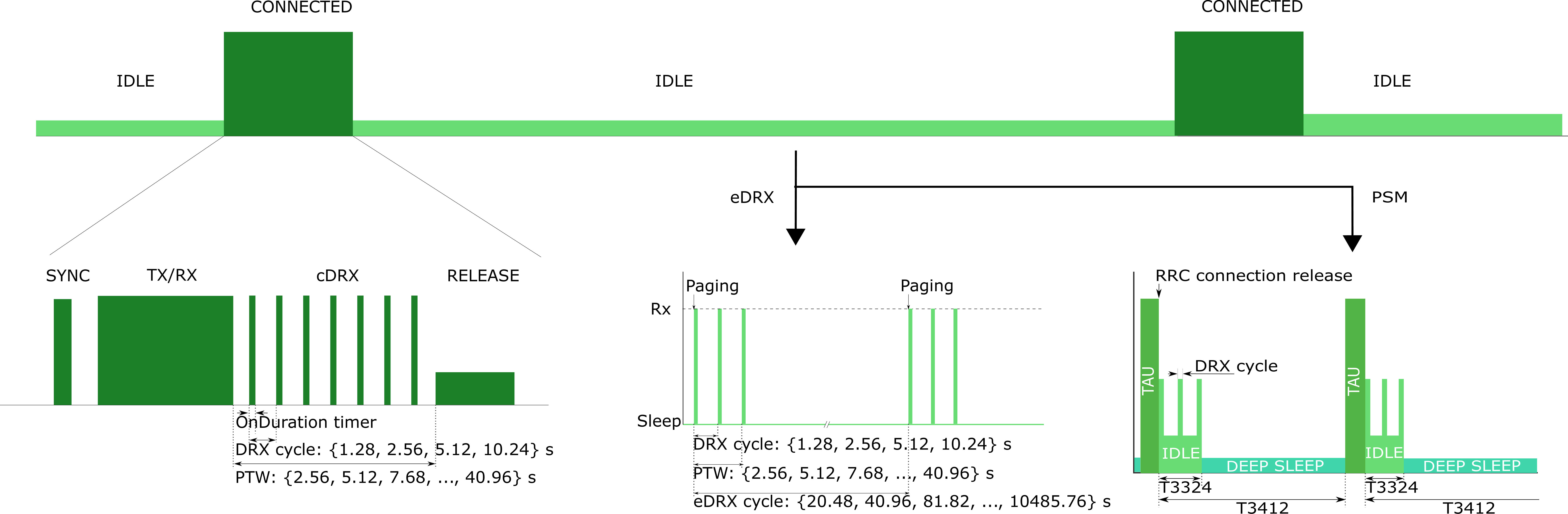}
    \caption{UE's main operational phases: Connected state, eDRX and PSM
      procedures in Idle state.}
    \label{fig:total}
  \end{subfigure}
  \hfill
  \begin{subfigure}[t]{0.325\textwidth}
    \includegraphics[width=\textwidth]{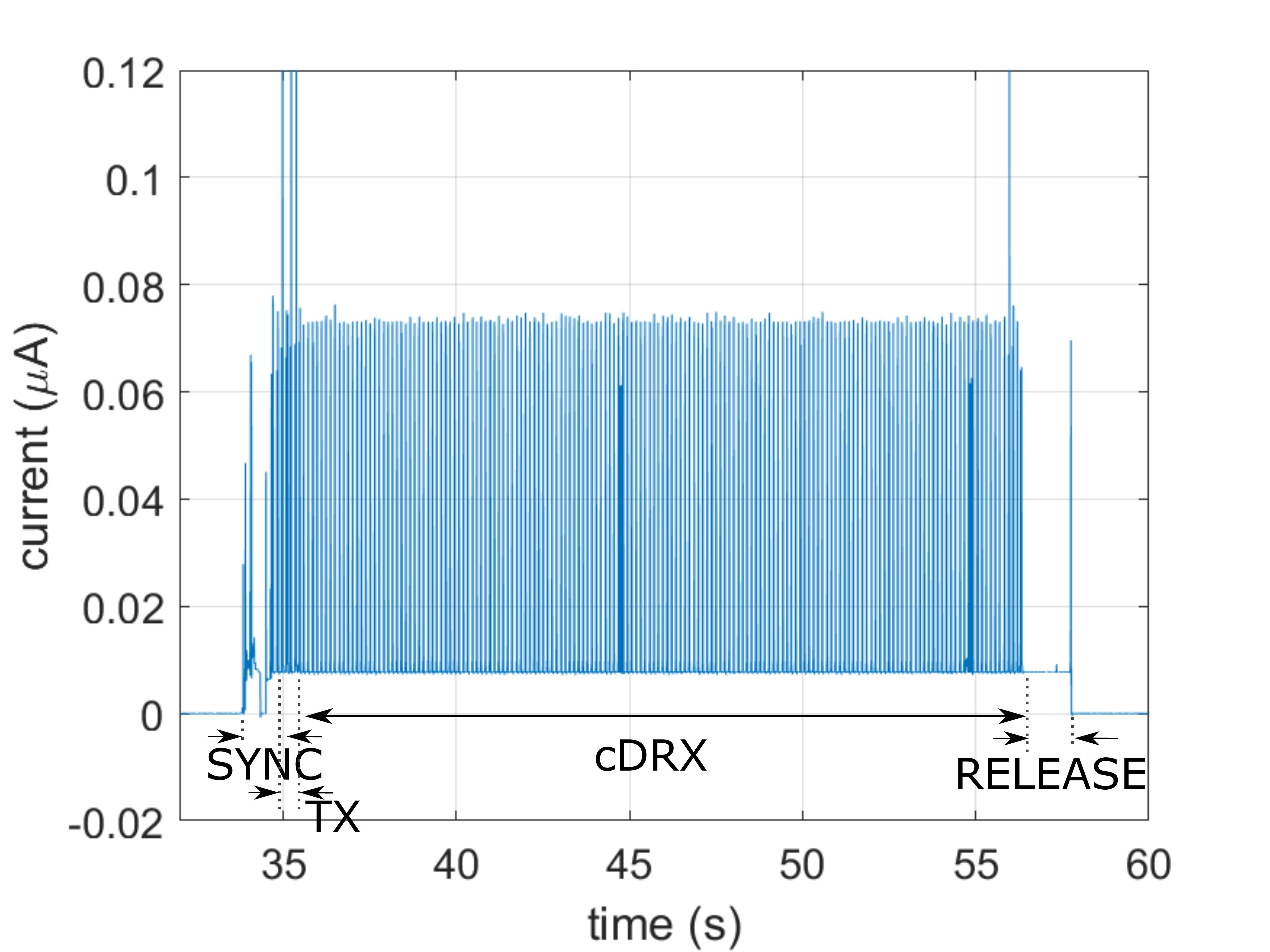}
    \caption{Current trace showing UE's Connected state.}
    \label{fig:conn}
  \end{subfigure}
  \hfill
  \begin{subfigure}[t]{0.325\textwidth}
    \includegraphics[width=\textwidth]{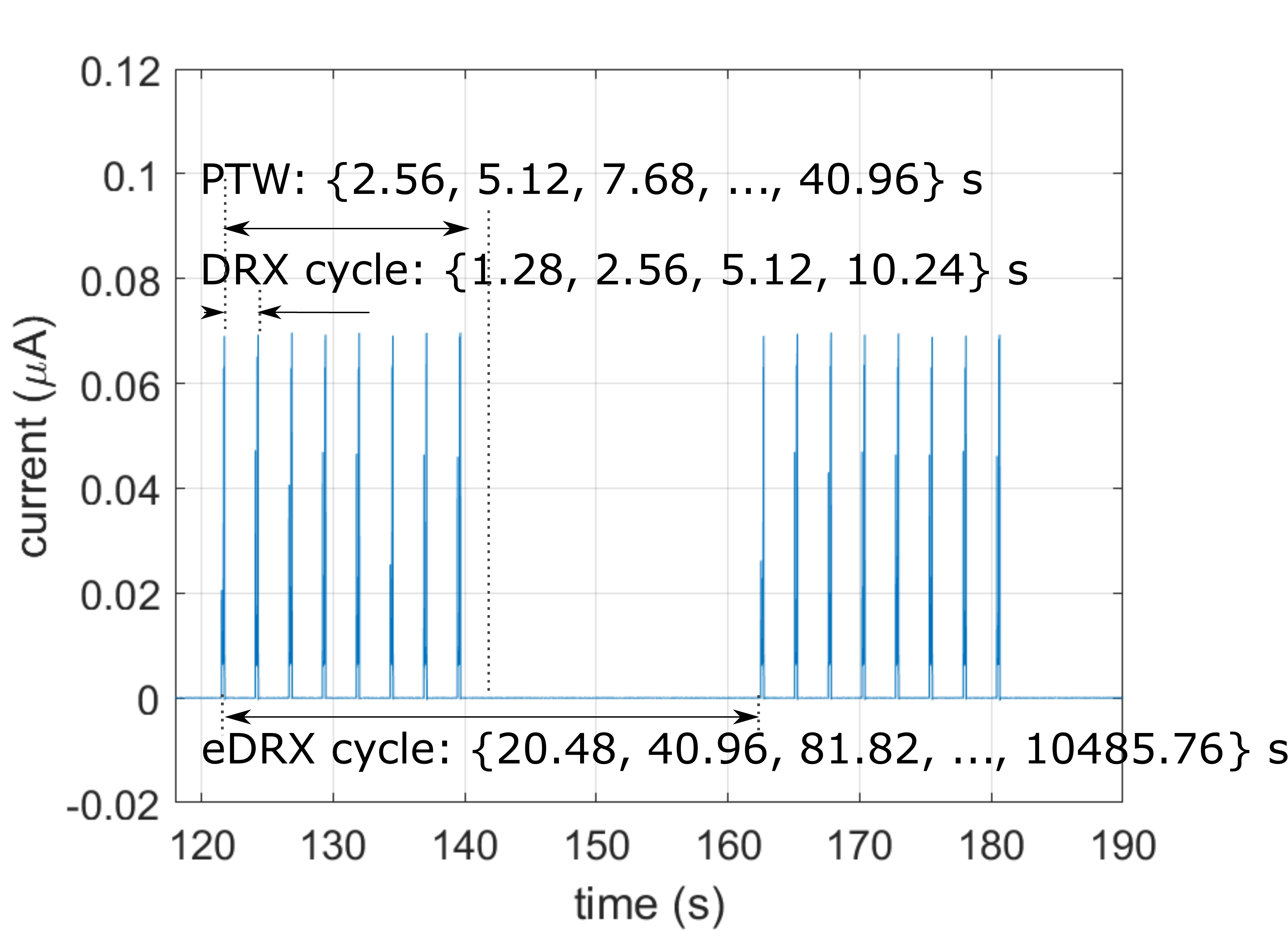}
    \caption{Current trace showing eDRX procedure.}
    \label{fig:edrx}
  \end{subfigure}
  \hfill
  \begin{subfigure}[t]{0.3255\textwidth}
    \includegraphics[width=\textwidth]{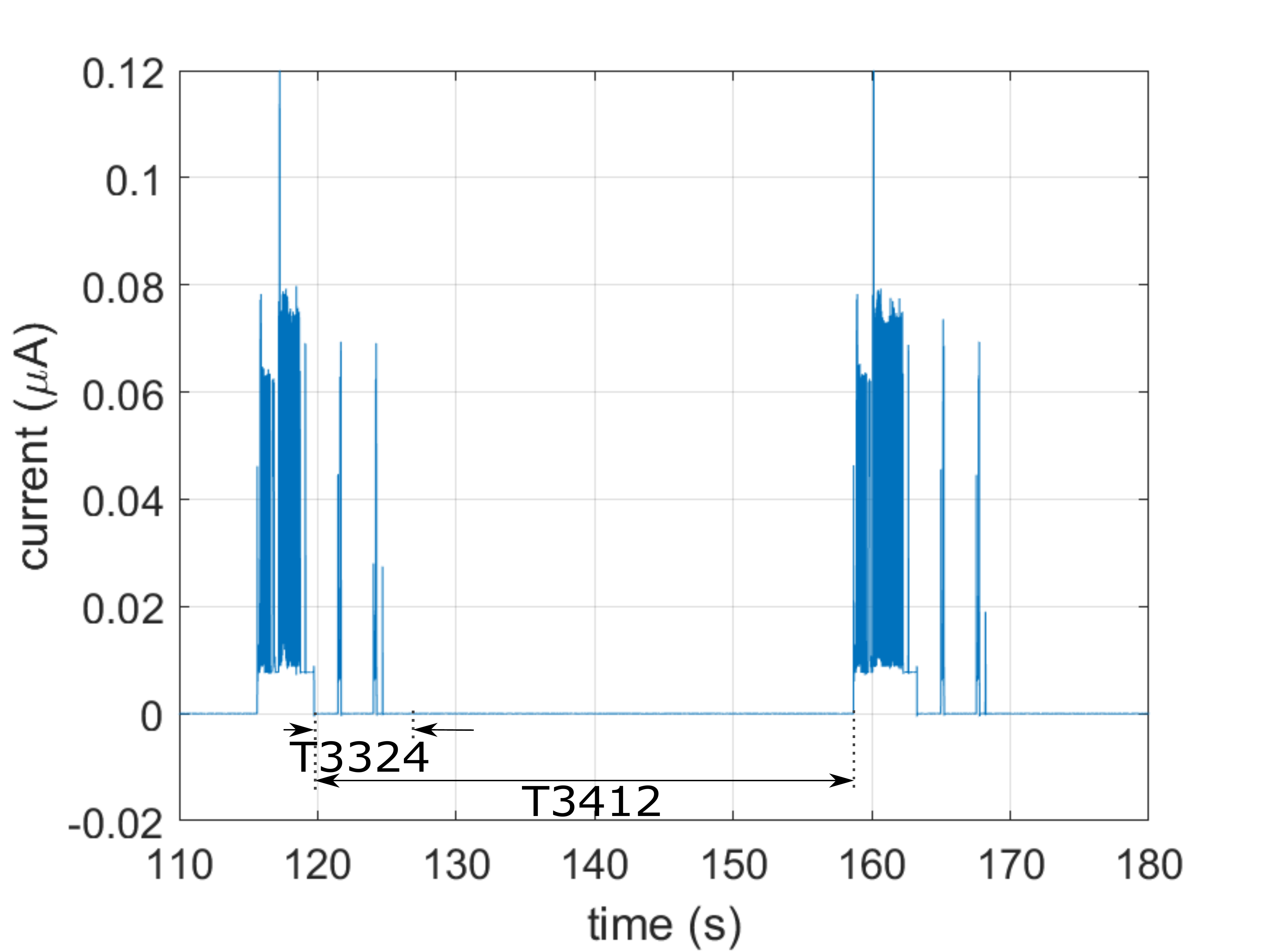}
    \caption{Current trace showing PSM procedure.}
    \label{fig:psm}
  \end{subfigure}
  \caption{UE's operational phases: illustrative scheme and empirical traces of
    current consumption.}

\end{figure*}

\begin{figure}
    \centering \includegraphics[width =0.8\linewidth]{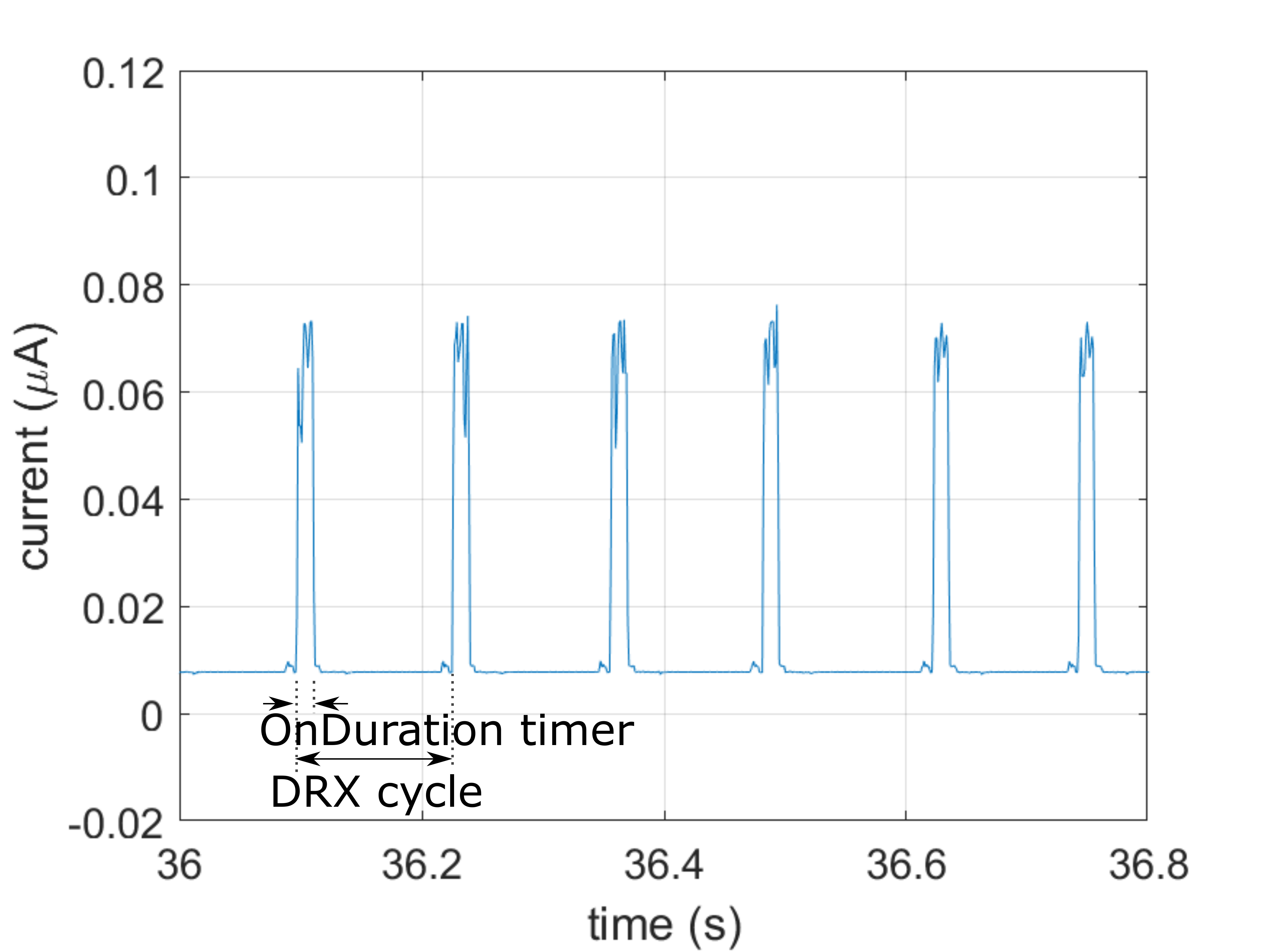}
    \caption{Example of the DRX procedure in Connected state.}
    \label{fig:cdrx}
\end{figure}

This section describes the operational phases of a \gls{ue} at different time
scales, as schematically illustrated in Fig.~\ref{fig:total}. At a macro time
scale, the \gls{ue} alternates between two main states: \textit{Connected} and
\textit{Idle}. In Connected state, the \gls{ue} maintains a control link with
the network. When such link is released, the \gls{ue} enters the Idle state. In
both states, the \gls{ue} periodically should check for the availability of
\gls{dl} messages at the \gls{bs}. To reduce power consumption, the \gls{ue} can
employ the \gls{drx} mechanism (see Fig.~\ref{fig:cdrx}), which consists in listening/sleep cycles whose
time duration is specified by the \textit{DRXCycle} parameter. The duration of
the listening period inside a \textit{DRXCycle} is specified by the
\textit{OnDurationTimer} and is expressed in multiples of $\sim$1 ms,
corresponding to the duration of a Paging Occasion, \ie a time interval during
which the \gls{ue} can receive notifications of pending packets from the
\gls{bs}.

The values of \textit{DRXCycle} and  \textit{OnDurationTimer} are set by the
\gls{bs}.

\begin{figure}
  \captionsetup[subfigure]{justification=centering}
    \begin{subfigure}[t]{\linewidth}
      \centering \includegraphics[width =0.8\textwidth]{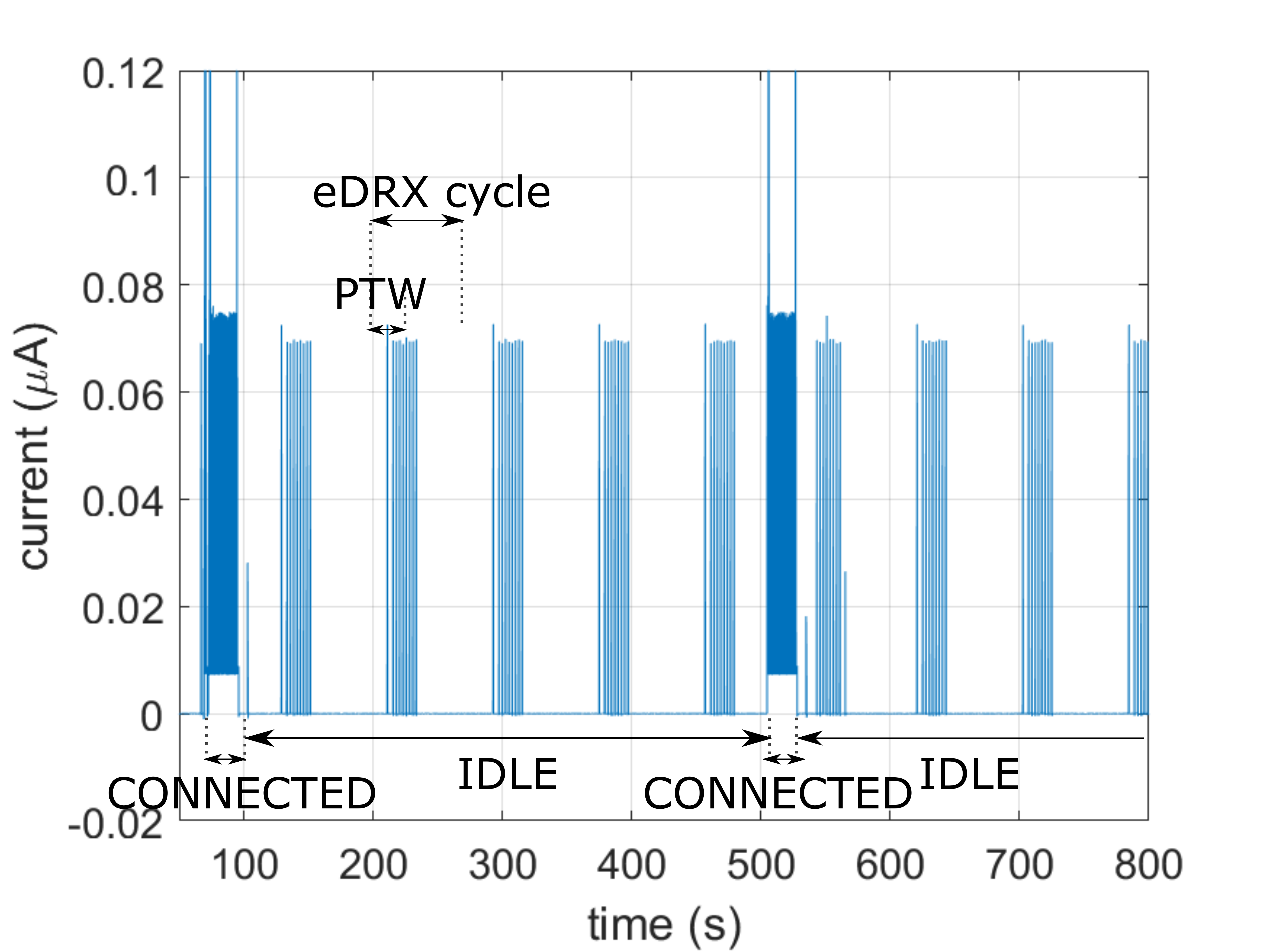}
      \caption{Experiment showing Connected state and eDRX
        procedure.}
      \label{fig:edrx_detail}
    \end{subfigure}

    \begin{subfigure}[t]{\linewidth}
      \centering \includegraphics[width =0.8\textwidth]{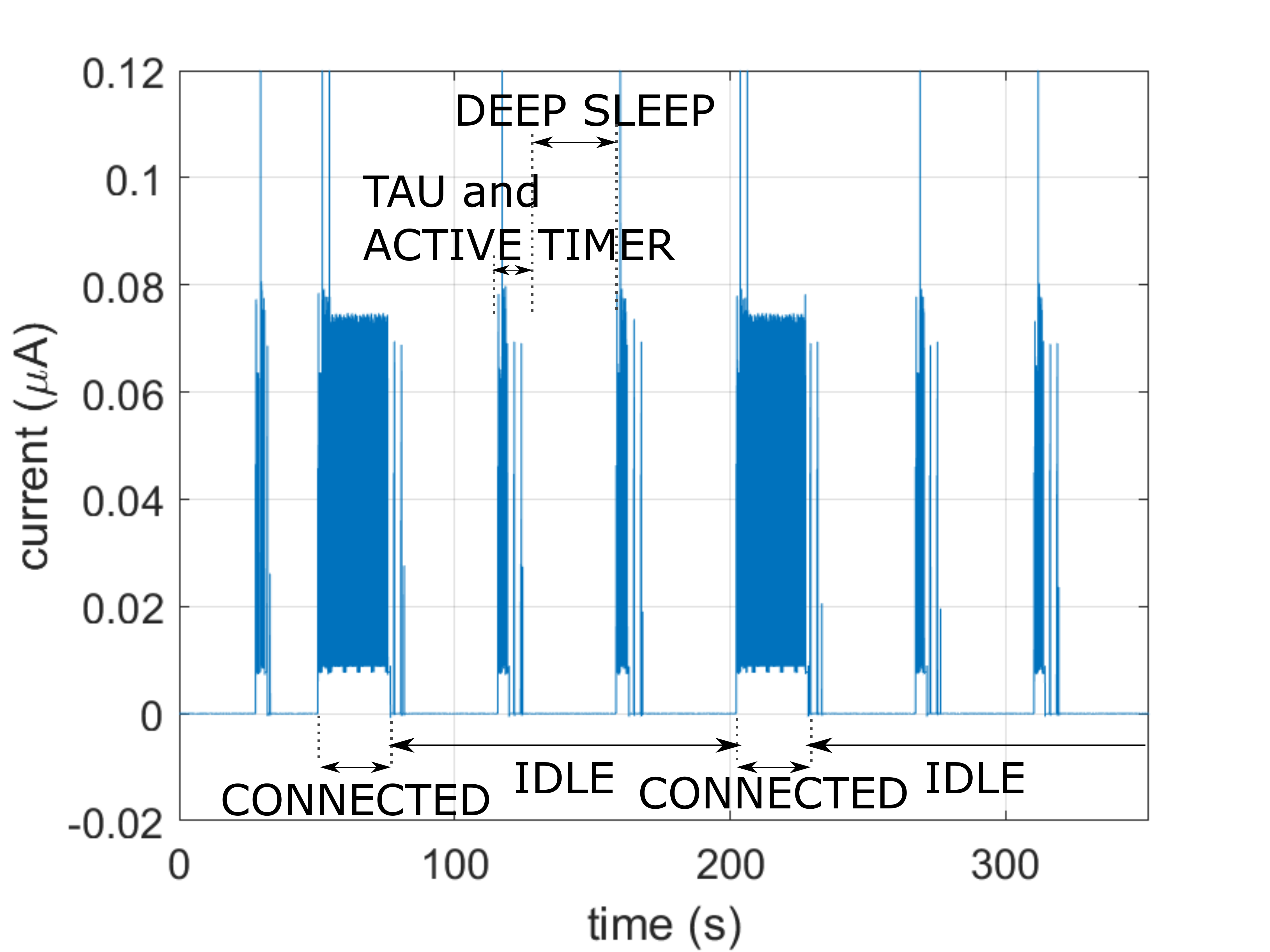}
      \caption{Experiment showing Connected state and PSM
        procedure.}
      \label{fig:psm_detail}
    \end{subfigure}
    \caption{Current traces of two experiments where the UE alternates between
      Connected and Idle states.}
    \label{fig:states}
\end{figure}

The plots in Figs.~\ref{fig:conn},~\ref{fig:edrx}, \ref{fig:psm}
and Fig.~\ref{fig:states} show some experimental current traces with periodic
\gls{ul} traffic. The periods of high current consumption correspond to
intervals in which the device is active transmitting, receiving or sensing the
channel for possible \gls{dl} messages. 
In the following, we examine in more detail the operations in Connected and Idle
states.

\subsubsection{Connected state}
\label{sec:phases-connected}
This actually consists in a combination of the following operations:
synchronization, transmission/reception, listening and release, which are described
below.

\noindent --\textit{Synchronization (SYNC):} this phase is performed by the
\gls{ue} to re\-/synchronize with the network whenever it exits from the Idle
state.
If the \gls{ue} does not have any allocated resources, it performs a random
access procedure to initiate the communication with the \gls{bs}.
In our experiments, we observe that this phase can have a variable duration.
\medskip

\begin{figure}
    \centering \includegraphics[width =0.8\linewidth]{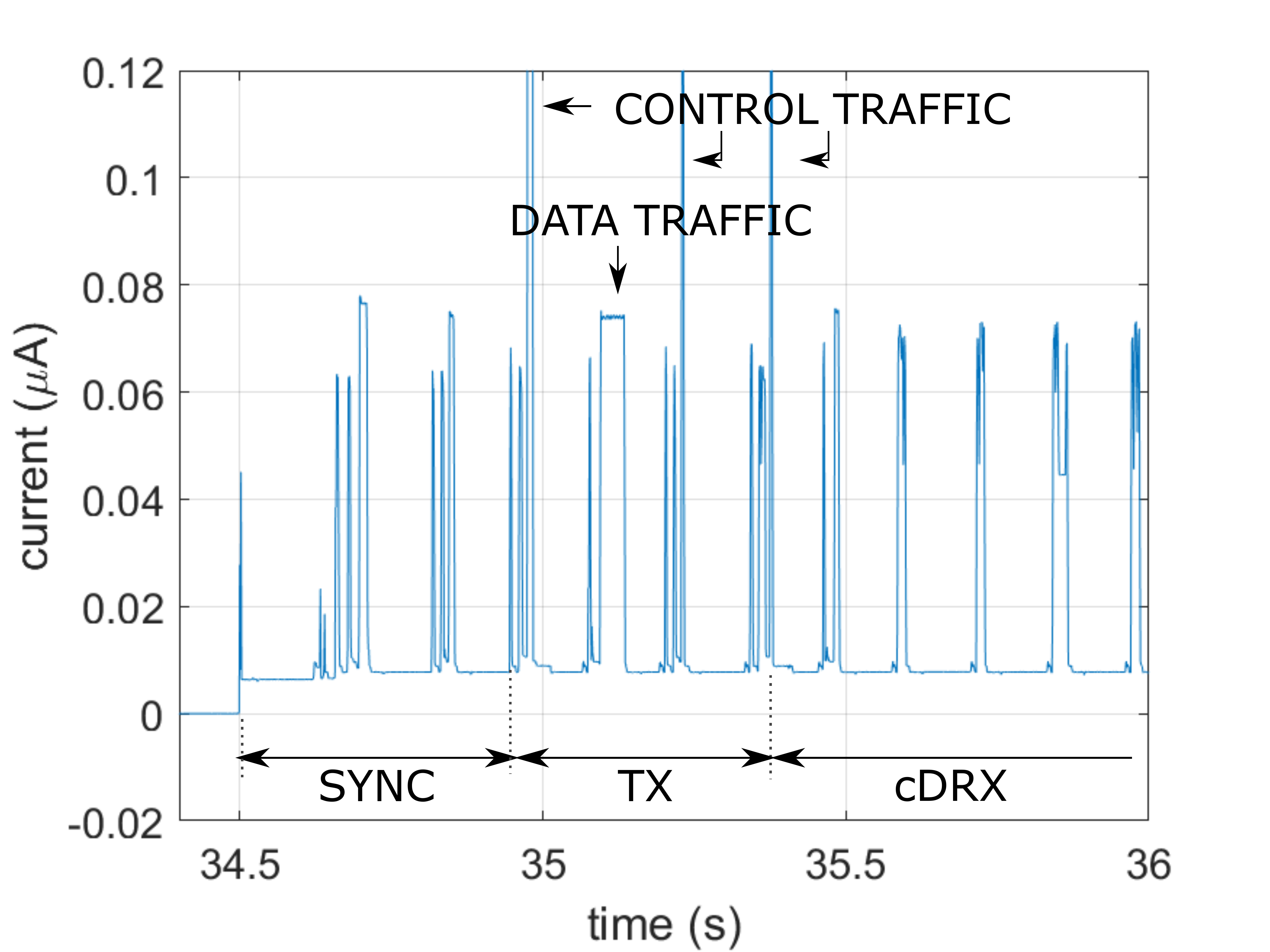}
    \caption{First part of the Connected state.}
    \label{fig:begin_conn}
\end{figure}

\noindent --\textit{Transmission and Reception (TX/RX):} this phase corresponds
to the transmission of (one or more) \gls{ul} messages, each followed by a
reception interval where the \gls{ue} waits for possible \gls{dl} data or
acknowledgement packets. In the current traces, the \gls{ul} transmissions are
preceded and followed by peaks of current consumption, as illustrated in
Fig.~\ref{fig:begin_conn}. Such peaks correspond to control signaling traffic.
The actual data transmission causes a lower peak, which lasts longer. In the
sequel, we consider as transmission phase the time between the highest peaks,
thus including signaling associated to the actual packet transmission.
As previously discussed, when the \gls{ue} exits the Idle state, the first TX/RX
phase is preceded by a SYNC phase to establish a control channel with the \gls{bs}.
The \gls{ue} then requests the allocation of transmission resources by
performing a Service Request operation, which in our analysis is considered part
of the TX/RX phase.
Instead, if the connection was only suspended rather than being released, the
Service Request is replaced by a Connection Resume procedure, which is lighter
in terms of control signaling.\medskip

\noindent --\textit{Listening Period:} In the Connected state the UE
maintains the so-called inactivity timer, which is restarted at any RX/TX event.
If the timer expires the UE performs the release operation and enters the Idle
state. The value of the timer is set by the base station, typically in the range
between 10 and 20 seconds. The \gls{ue} can ask the network to set the value of
this timer to zero using the \gls{rai} flag, explained in
Sec.~\ref{subsec:ecl-rai}. In this case, the UE leaves the Connected state
immediately after a TX/RX event.

While the inactivity timer is counting down, the \gls{ue} might keep the radio
on, always listening, or perform \gls{cdrx}. During \gls{cdrx}, the \gls{ue}
alternates between high energy periods of listening for scheduling information
and low energy periods of sleeping. During sleeping periods, the radio consumes
$\sim$90\% less energy.
In case of available \gls{dl} messages, the \gls{ue} can directly perform a
TX/RX without any SYNC operation. An example of the current consumption in
\gls{cdrx} is given in Fig.~\ref{fig:cdrx}. \medskip

\noindent --\textit{Release:} the \gls{ue} releases the connection with the
\gls{bs} and leaves the Connected state entering the Idle state.\medskip

\subsubsection{Idle state}

\begin{table*}
     \centering
     \caption{Main timers for DRX, eDRX and PSM. Note that timers used in DRX
       also apply to cDRX and eDRX.}
     \label{tab:timers}
     \begin{tabular}{p{1cm}p{2.5cm}p{8.8cm}p{1.5cm}p{1.5cm}}
       \toprule
       Mode & Timer & Description & Min Value & Max Value \\
       \midrule
       \multirow{2}{*}{DRX}& OnDurationTimer & Time spent in active listening & 1 ms & 200 ms \\
            & DRXcycle & Time interval between the beginning of two active listenings & 2 ms & 2.56 s \\
       \midrule
       \multirow{2}{*}{eDRX} & PTW & Duration of Paging Occasions monitoring, composed of multiple \gls{drx} cycles & 2.56 s & 40.96 s \\
            & eDRXcycle & Time interval between the beginning of two PTWs & 20.48 s & 10485.76 s \\
       \midrule
       \multirow{2}{*}{PSM} & T3324 & Active Timer: duration of DRX/eDRX within Idle state (listening for paging) & 2 s & 410 hours\\
            & T3412 & TAU timer: interval between two TAUs & 2 s & 410 hours \\
       \midrule
            & Inactivity timer & Time spent in Connected state, after the end of the last TX/RX & 0 s & 65.536 s \\ 
       \bottomrule
     \end{tabular}

  \end{table*}

  In the Idle state, the \gls{ue} may utilize two power saving mechanisms, in
  addition to normal \gls{drx}: \gls{edrx} or \gls{psm}. These mechanisms,
  better described below, are based on timers that are negotiated with the
  network (see Table~\ref{tab:timers}). \medskip

\noindent --\textit{\gls{edrx}:} this mechanism is similar to \acrshort{cdrx},
but with more sporadic listening periods. An \gls{edrx} cycle, indeed,
corresponds to a sequence of \gls{drx} listening/sleep cycles, called \gls{ptw},
followed by a long sleep period (see Figs.~\ref{fig:total},~\ref{fig:edrx},~\ref{fig:edrx_detail} for reference). The overall duration of an \gls{edrx} cycle is
determined by the \textit{eDRXcycle} parameter, while the period in which the
\gls{ue} performs the \gls{drx} cycles is determined by the \textit{\gls{ptw}}
parameter~\cite{3gpp136304}. Therefore, the time difference between
\textit{eDRXCycle} and \textit{PTW} gives the duration of the sleep period in an
\gls{edrx} cycle. When not listening, the radio is
off. \medskip

\noindent -- \textit{\gls{psm}:} this is the most effective power saving
technique supported by \gls{nbiot}. During \gls{psm}, the \gls{ue} switches off
its radio for a long period (deep sleep), but keeps the registration to network:
therefore, when exiting \gls{psm}, the \gls{ue} just needs to perform a
Connection Resume operation. Moreover, in \gls{psm} the \gls{ue} periodically
performs a \gls{tau} operation to communicate its location to the network. The
\gls{psm} is characterized by two timers, namely T3412 and T3324 (see Figs.~\ref{fig:total},~\ref{fig:psm},~\ref{fig:psm_detail} for reference). The T3412
timer (or \textit{TAUTimer}) defines the time interval between two \gls{tau}
operations. Each \gls{tau} is followed by a period, whose duration is defined by
the timer T3324 (\textit{ActiveTimer}), during which the \gls{ue} listens for
paging, similarly to what happens during the \gls{ptw}. After this time, the
\gls{ue} enters into a deep sleep state and is not longer reachable by the
network. The \gls{ue} exits the sleep state when T3412 expires or when a new
\gls{ul} data becomes available. Fig.~\ref{fig:psm} shows the current
consumption for the \gls{tau} operation, followed by the listening for paging
interval, whose duration is determined by the T3324 timer.\medskip

We observe that  the TAU timer in \gls{psm} can be almost 17 days long.
Therefore, a device entering in \gls{psm} will consume a minimum amount of
energy, but may be unreachable from the network for several days, if no
\gls{ul} transmission is required.
On the contrary, adopting the \gls{edrx} power saving mechanism, the \gls{ue}
can be contacted by the network within a limited time interval, but at the cost
of higher energy consumption.

\subsection{\acrshort{ecl} and \acrshort{rai}}
\label{subsec:ecl-rai}
One of the objectives of \gls{nbiot} is providing reliable communication to
devices in harsh conditions, such as parking garages and ground pits. Therefore,
the \textit{\gls{ecl}} feature is introduced to tune the robustness of the communication.
Robustness is primarily achieved by repeating the messages up to thousands of
times, at the cost of a reduced data rate and an increased delay and energy
consumption. The \gls{bs} can set the \gls{ecl} parameter based on the received
\gls{nrsrp}, a metric indicating the power of the \gls{lte} reference signals.
The \gls{3gpp} identifies three different coverage levels, namely
\textit{Normal} (\eclZero), \textit{Robust} (\eclOne) and \textit{Extreme} (\eclTwo),
which are defined in terms of the target \gls{mcl}, which is set to 144, 154,
and 164 dB for the three levels, respectively.\footnote{\gls{mcl} is the largest
  attenuation between the transmitter and the receiver that can be supported by
  the system with a defined level of service.}
Each level is associated to a certain setting of some transmission parameters,
including the transmit power, the subset of subcarriers, the number of
repetitions of random channel access, and the maximum number of transmission
attempts. These result in prolonged transmission and reception under challenging
conditions. In the worst case, in \eclTwo, the number of repetitions may reach
2048 and the transmission delay 10 seconds.
The thresholds for each \gls{ecl} class and the associated transmission
parameters are determined by the operators. The \gls{bs} monitors the signal
strength of a target device on both the uplink and what the device reports for
the downlink and decides its ECL level. The device does not have any control on
the \gls{ecl} parameter, but in our experiments it was possible to retrieve its
current value by using appropriate diagnostic commands.

Conversely, the \gls{ue} can control the \textit{\acrfull{rai}} flag that is
carried into signaling messages before any \gls{ul} transmissions. This flag is
used to notify the \gls{bs} that, after the upcoming \gls{ul} transmission, the
\gls{ue} is expecting: (i) another \gls{ul} transmission; (ii) a \gls{dl}
message; (iii) none of the previous. Based on this signalling, the \gls{bs} can
release the connection beforehand (see Table~\ref{tab:rai}), so that the \gls{ue}
can reduce the time spent in \gls{cdrx} phase, awaiting incoming \gls{dl}
transmissions.
The effects of this parameter will be explored in the following sections.

  \begin{table}
    \centering
    \small
        \caption{Possible values of \acrfull{rai} field~\cite{sara-n2-man}.}
    \label{tab:rai}

    \begin{tabular}{p{1.8cm}p{5.5cm}}
    	\toprule
    	Flag value & Meaning \\
  		\midrule
      0x000 & no flags set: remain in the Connected state for the duration of
              T3324 timer \\
      0x200 & RAI:  release after next \gls{ul} message \\
	    0x400 & RAI:  release after next \gls{ul} message has been replied to \\
	    \bottomrule
    \end{tabular}
        \vspace{-0.7cm}
  \end{table}

  \noindent{\bf Takeaways.} From this quick introduction to \gls{nbiot}
  operations it is apparent that a proper tuning of the parameters of the power
  saving mechanisms is crucial to control the trade-off between maximum latency
  and energy consumption. Moreover, operators can control the robustness of the
  connection by choosing the transmission parameter settings for each of the
  three different coverage levels entailed by the standard. However, the
  sensitivity of such adjustments is still largely unknown. Shedding light on
  these aspects is one of the goals of this study.

%% file: methodology.tex
\section{Methodology}
\label{sec:methodology}

In the following, we discuss our experimental setup, motivating our choices with
respect to: (i) experimental boards, (ii) tools for measuring energy
consumption, (iii) measurement setup and (iv) collection of metadata for
contextualizing the measured performance.

\subsection{Experimental setup}
\label{sec:exp_set}

\noindent{\bf Experimental boards (\gls{ue}).}
During the measurement period, both operators deployed NB-IoT using 15 KHz
single-tone over band B20 (800 MHz) in Guardband.
We have used two, compatible with this configuration, off-the-shelf
NB-IoT modules, namely u-blox SARA-N211-02B~\cite{datasheet:saran211} and
Quectel BC95-G~\cite{datasheet:bc95g}.
These modules are among the first commercially available LTE Cat NB1 \glspl{ue}
and they have been certified by a number of mobile operators.
The first module supports data rates up to 27.2~kbps in \gls{dl} and
31.25~kbps in \gls{ul}. 
Quectel BC95, when operating in single-tone, supports up to 25.2~kbps in
\gls{dl} and 15.625~kbps in \gls{ul}. 
Since, the form factor of these modules does not lend itself to
experimentation, they are sold as a part of a development board (\ie dev-kit)
that facilitates powering the module and interfacing with it via USB.

\noindent{\bf Measuring energy consumption.} We have employed the Otii Arc power
measurement device for tracking energy
consumption.\footnote{https://www.qoitech.com/products/standard}
This device
can be used as both a power supply unit for the tested IoT device and a current
and voltage measurement unit. It provides up to 5 V with a high resolution
current measurement with a sampling rate up to 4000 samples per second in the
range from 1 $\mu A$ to 5 $A$. To characterize the energy consumption associated
with different \gls{nbiot} operations,
we need to ensure that the meter measurements correspond to the current drawn by
the module only, and not that drawn by the entire dev-kit. When using
SARA-N211-02B, this can be obtained by powering the module directly with the
Otii Arc power measurement device. 
Quectel BC95 does not readily allow
for a similar setup. In this case, we had to remove three resistors from the
dev-kit and solder a zero-ohm resistor on the power path to isolate the module
power supply from the dev-kit.\footnote{A zero-ohm resistor acts simply as a
  jumper or a wire.}

\begin{figure}[]
    \centering
\includegraphics[width=0.9\linewidth]{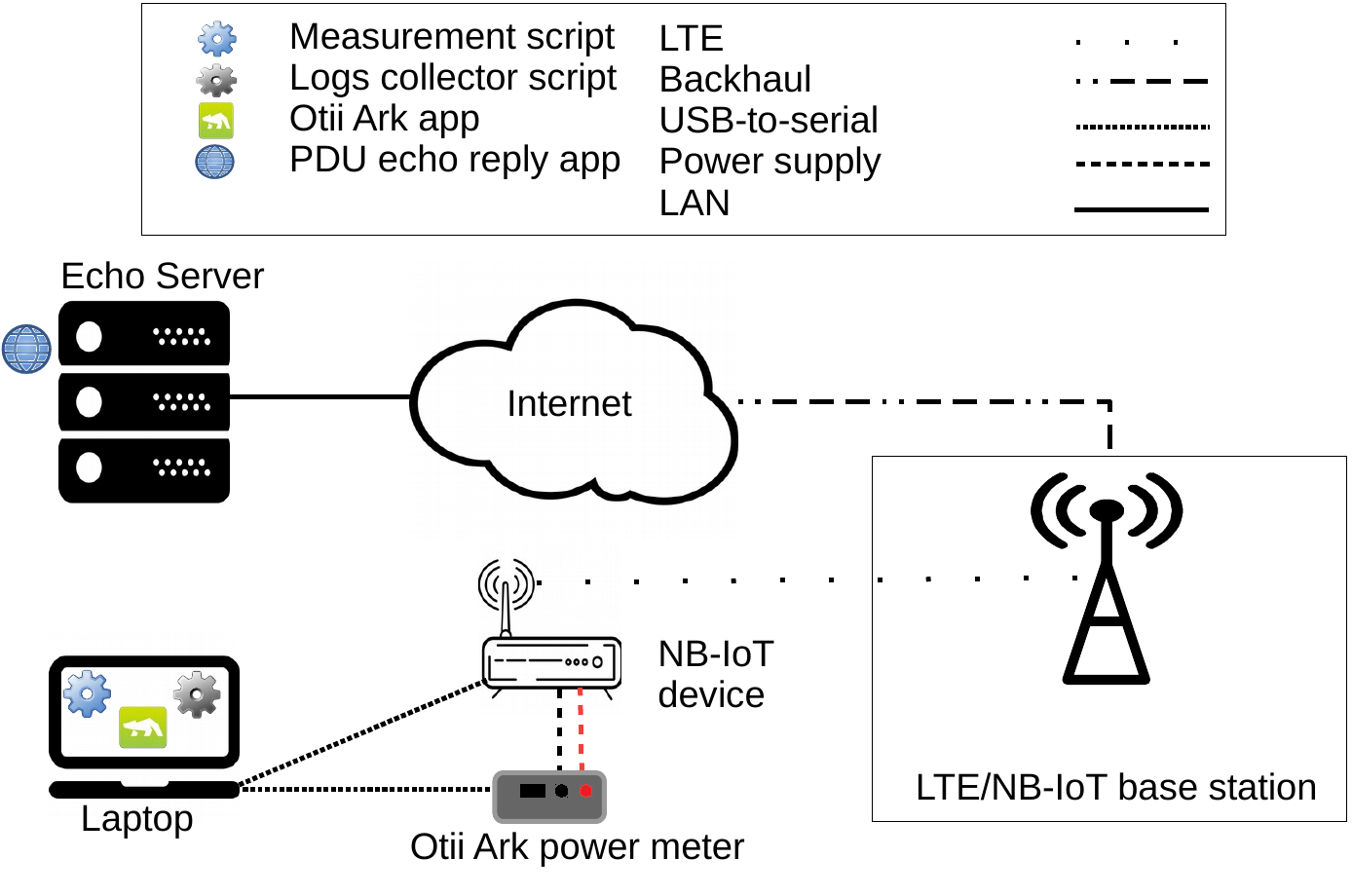}
\caption{Experiment setup.}
\label{fig:experiment_setup}
\vspace{-2em}
\end{figure}

\noindent{\bf Measurements.} We have connected each dev-kit to a laptop, where
we run a set of scripts to manage the \gls{nbiot} \gls{ue}'s authentication,
registration to the network, and \gls{rrc} configuration. The \gls{nbiot}
modules use commercial subscriptions to connect to two major mobile operators in
a European country. Both operators deploy \gls{nbiot} in guard band, which
reduces the likelihood of interference with LTE. To measure power consumption,
we send UDP packets of various sizes (12, 20, 128, 256 and 512
bytes) 
to a well-provisioned server that echoes them back. The packets are sent at
different frequencies depending on the experiment.
Fig.~\ref{fig:experiment_setup} shows our experiment setup. Our goal is
measuring a baseline performance, thus we avoid generating traffic through
applications (\eg MQTT, CoAP), as this would add the complexity of the
application on top of an already complex setup.
We have repeated the measurements under various power
management configurations, which we describe
below. 
Further, we have run the experiments at various locations, which we then group
based on their coverage condition into ``Good coverage'' and ``Bad coverage''.
We create poor coverage conditions in two ways: 1) by
using signal attenuators and 2) by placing the modules in a specially designed
metallic box. 
This setup allows for repeatability of experiments. For some of the
experiments, we used a different method to simulate poor coverage in a real life
scenario: we placed the modules in a deep basement, in a similar fashion to a
metering device use case. The performance at the basement is similar to the
performance when using the attenuators and the special box. In these bad
conditions, normal LTE mobile devices are out of coverage.
Fig.~\ref{fig:metadataVrsrp} presents the \gls{rsrp} values of each group of
locations, which can be used as a guide to reproduce our experiments.

\noindent{\bf Data collection}
We use the same laptop to control both the dev-kit and the Otii Arc power
measurement device.
Besides measuring power consumption, we also track \gls{rtt},
packet loss and throughput.
We used a set of \texttt{AT} commands for collecting connection metadata.
These include \gls{rrc} Connection and Release events, \gls{snr},
\gls{txpower}, \gls{ecl}, \gls{pci}, \gls{rsrp} and \gls{rsrq}.
In addition, we use a software called UEMonitor, developed by Quectel, to
collect and decode debug messages generated by the \glspl{ue}, as well as NB-IoT
control plane messages such as the \gls{dci} messages.

Our measurements are spread over several months between October 2018 and October
2019, which gives us the opportunity to track the maturing of the measured
deployments.
Overall, we have sent about 13000 packets, which corresponds to 9 days at the
rate of one packet per minute.
70\% of these experiments were run using the default settings and 30\% using the
\gls{rai} flag.
Furthermore, 75\% of the experiments were conducted in good coverage conditions.
Also, one third of the experiments involved sending a 20-byte UDP packet, the
remaining two thirds were split among packet sizes of 12, 128, 256 and 512
bytes.

\subsection{Experiments}
\label{sec:exp}

We focus on three operation modes / scenarios corresponding to the possible
setting of the \gls{rai} flag (see Table~\ref{tab:rai}):
\begin{enumerate}
\item {\em TX/RX default timers (\gls{rai}-000)}:  The \gls{ue} sends an \gls{ul}
packet to a remote server, which echoes it back.
During this it sticks to the default setting, in which
it remains in the Connected state, monitoring the channel for paging messages
after an \gls{ul} transmission for the duration of the inactivity timer. Then,
after the \gls{rrc} Release, it enters the \gls{psm}.
This scenario corresponds to applications that require two-way communication,
\eg reliable monitoring or alarm services.
\item{{\em TX/RX and release (RAI-400)}: Here, the \gls{rrc} connection is
    released once the response from the server is received. The application
    scenario is again a two-way communication service. The immediate release is
    intended for optimizing energy consumption.}
\item{{\em TX and release (\gls{rai}-200)}: In this case, the \gls{rrc} connection is
released after sending the \gls{ul} packet (i.e. the reception of the echo
packet is skipped).
This corresponds to services without strict reliability requirements.}
\end{enumerate}

Recall that each of these scenarios comprises two distinct states: Connected and
Idle, as described in Sec.~\ref{sec:nbiot_primer}.
We examine the energy consumption during the Connected and Idle state,
separately. To do this, we need to identify which state and phase the device is
in at any particular time point.
We present our algorithm for automatically identifying the device state from the
experiment logs in Appendix~\ref{sec:dataPreprocessingAndMetadata}.

%% file: metadata_quality.tex
\section{metadata as a proxy for performance}
\label{sec:metadata_quality}



\begin{figure*}
  \centering
  \begin{subfigure}[t]{0.49\linewidth}
    \centering
    \includegraphics[width=1.125\linewidth]{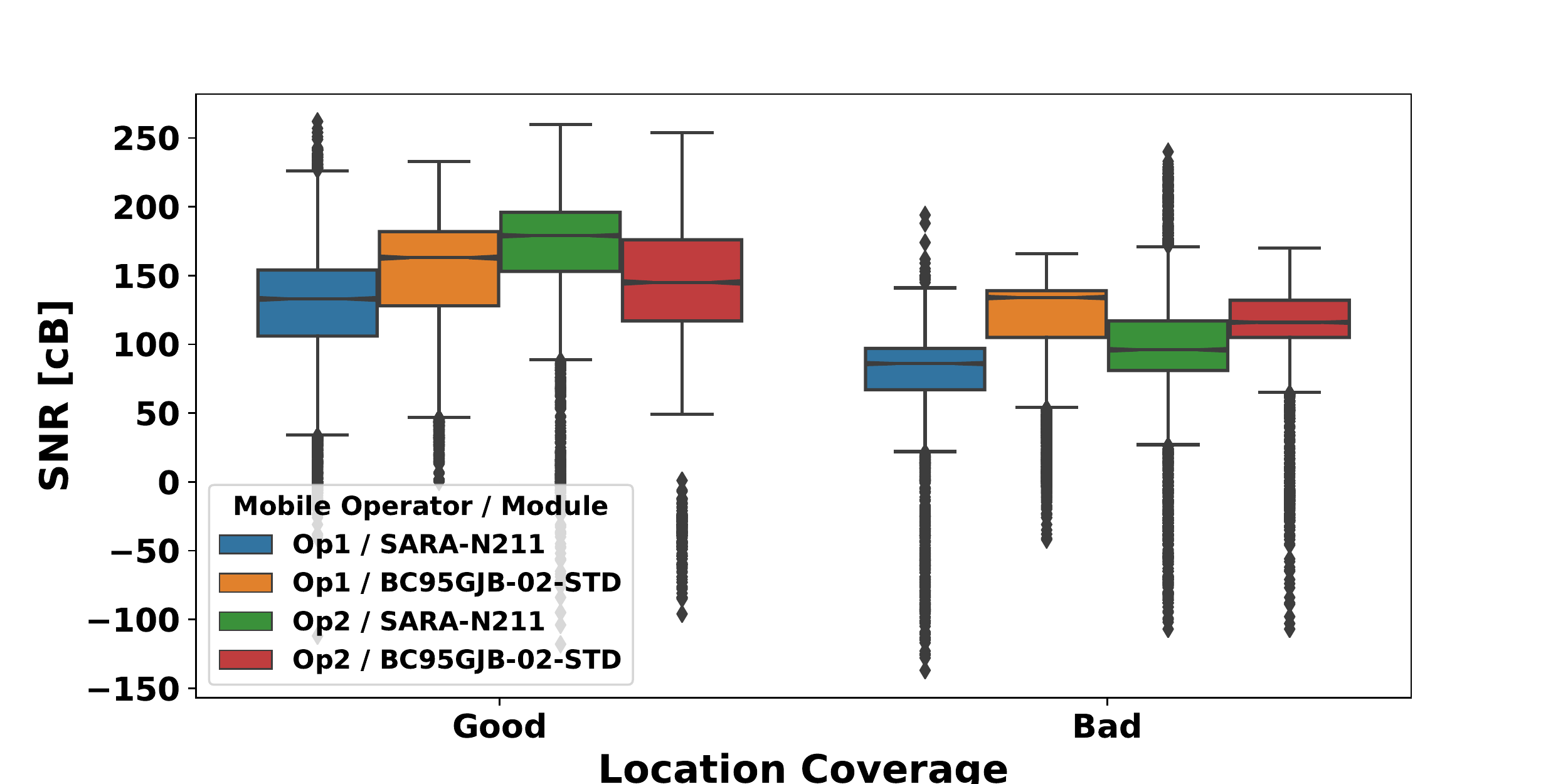}
    \caption{\textbf{SNR}}
    \label{fig:metadataVsnr}
  \end{subfigure}
  \begin{subfigure}[t]{0.49\linewidth}
  \centering
  \includegraphics[width=1.125\linewidth]{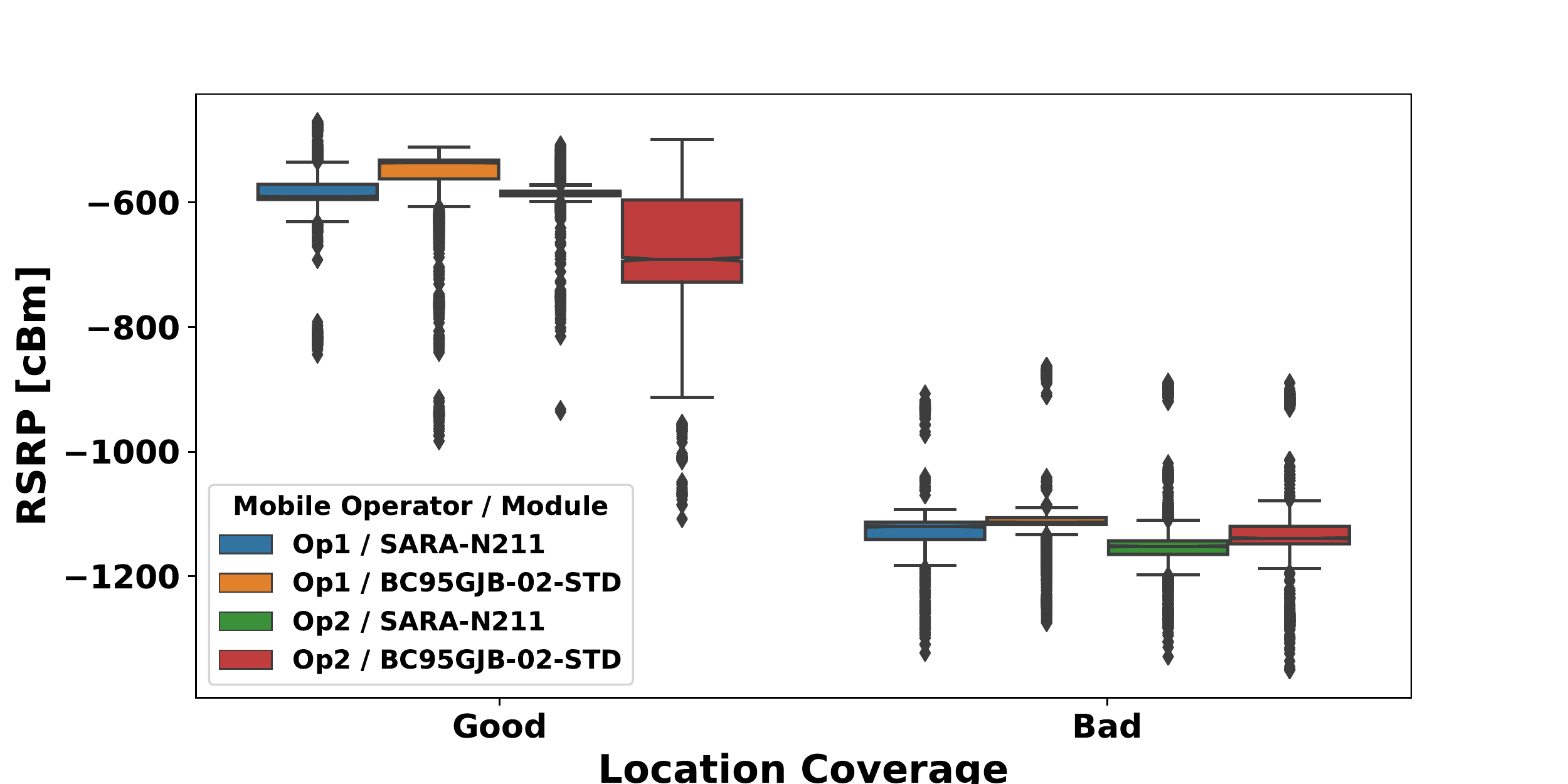}
  \caption{\textbf{RSRP}}
  \label{fig:metadataVrsrp}
  \end{subfigure}
  \caption{Distributions of metadata in locations with weak signal (bad
  coverage) and good signal (good coverage).}
  \label{fig:metadataVlocation}
  \vspace{-1em}
\end{figure*}

It is important to collect accurate and frequent metadata,
as they are an indication of performance and help diagnose
problems.
Both devices report metadata through AT command requests.
These requests consume energy (around 15 mJ in our measurements
) and may take several seconds to fulfill.
The response time increases with worsening signal conditions.
For instance at locations
with very bad coverage the request may time out and some
of the metadata might not be reported or have obviously wrong values (\eg
\gls{snr} value of -30000).
In this Section, we examine the metadata reporting accuracy and investigate
which metadata metrics better reflect network and energy performance, so that
users can get the most value out of this costly operation.
Both devices report power ratios in cB ($1 dB = 10 cB$) and power in cBm
($1 dBm = 10 cBm$).
We start by comparing the behaviour of the most commonly used metadata metrics:
\gls{snr} and \gls{rsrp}.
Fig.~\ref{fig:metadataVlocation}, presents the distributions of \gls{snr} and
\gls{rsrp} when
we group the measurements based on the expected signal quality of the
measurement location.
As we will present in the sequel, the biggest effect on performance is caused
by the choice of operator and module, thus in this Figure and the rest of the
paper, we control our measurements for these two variables.
The \gls{snr} distributions are very wide, with a significant overlap between
the good and the bad locations.
Further, the median values between the two locations show a small difference
between 30 and 80 cB.
In contrast, the \gls{rsrp} distributions better reflect the signal quality at
each location, there is significantly less overlap in the distributions and the
distributions are also narrower.

\begin{figure*}
  \centering \includegraphics[trim=0 0 0 40, clip, width=1\textwidth]{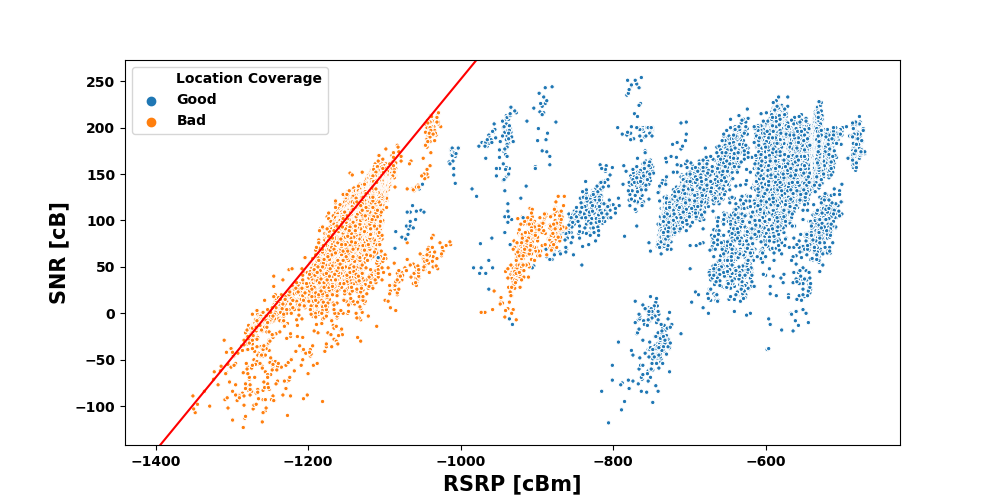}
  \caption{Mapping of \gls{snr} and \gls{rsrp}.
  The red line represents the ideal mapping assuming no interference (applies
  only to bad coverage data).}
  \label{fig:snrVrsrp}
  \vspace{-2em}
\end{figure*}


\gls{sinr} and \gls{rsrp} are connected by
$SINR=\frac{S}{I+N}=\frac{12*RSRP}{I_{tot}+N_{tot}}$~\cite{landre2013lte},
where we assume that
\gls{rsrp} is free from noise and interference and includes only useful
(reference signal) power.
$I_{tot}$ and $N_{tot}$ are the interference and noise computed over the whole
180 kHz bandwith and since \gls{rsrp} is the power of a single 15 kHz subcarrier
we multiply it by 12, which is the number of subcarriers.
Both operators deploy NB-IoT in the guard band, thus there should be no
interference from normal LTE traffic.
Also, the measurements were performed soon after the NB-IoT was deployed, so the
number of other users is very small, minimizing interference from
neighboring NB-IoT cells.
Thus, the main component of the denominator is noise, which is affected by
temperature and the noise figure of the receiver, so we expect the noise to
not fluctuate much.
Under these assumptions, \gls{rsrp} and \gls{snr} should have a linear
relationship when expressed in cB.
Fig.~\ref{fig:snrVrsrp} shows the connection between \gls{rsrp} and \gls{snr}.
The red line is the ideal mapping of \gls{rsrp} values to \gls{snr} under the
assumption that there is no interference, for typical values of thermal noise
density and receiver noise figure, $N_{thermal}=-1740 cBm/Hz$ and
$NF_{receiver}=70cB$, respectively: $SNR_{cB}=RSRP_{cBm}+1252$
(proof in Appendix~\ref{sec:snr_rsrp_mapping}).
This relationship is verified for the bad coverage measurements, but not for
the good coverage measurements.


We briefly report our observations for the rest of the metadata.
Both modules log the following metadata: \gls{rssi}, \gls{snr},
\gls{rsrp}, \gls{rsrq}, \gls{ecl} and \gls{txpower}.
\gls{rssi} values have similar distributions to the \gls{rsrp} values and
are typically between -470 and -600 cBm in good locations and around -1030 cBm
in bad.
Thus, they are typically 60 cBm higher than \gls{rsrp} in good and around 100
cBm higher than \gls{rsrp} in bad conditions.
The \gls{rsrq} values are around -108 cBm for good conditions and slightly worse
between -108 and -113 cBm for bad conditions.
\gls{rsrq} has very small variation across different conditions, making it
poorly correlated with performance.
Since \gls{rssi} does not provide further information over \gls{rsrp}, we can
safely disregard it.


Only the \gls{rsrp} and \gls{rsrq} are reported to the eNodeB and
from these the eNodeB can estimate the \gls{rssi}~\cite{3gpp36214}.
The \gls{rsrq} measurement provides additional information when \gls{rsrp} is
not sufficient to make a reliable handover or cell reselection decision.
In contrast, \gls{rsrp} is the most important metadata metric.
During the \gls{rach}, the \gls{ue} sets its \gls{ecl} and \gls{txpower}
based on the \gls{rsrp} thresholds it receives from the eNodeB.
If the \gls{ue} is unable to connect, it increases its \gls{txpower}
by 2 dB increments, until it achieves connectivity or until it reaches a
predefined number of preamble transmission attempts per \gls{ecl}
supported in the serving cell.
Then, it increases its \gls{ecl} by 1 and sets the \gls{txpower} to maximum and
repeats the process.
The \gls{rsrp} thresholds and the number of transmission attempts per \gls{ecl}
are set by the operator~\cite{liberg2019cellular}. 
For the above reasons, we will focus mostly on \gls{rsrp} in the sequel, as it
is the metric the reflects best performance and energy consumption.



\begin{figure}[t]
  \centering\includegraphics[width=1.1\linewidth]{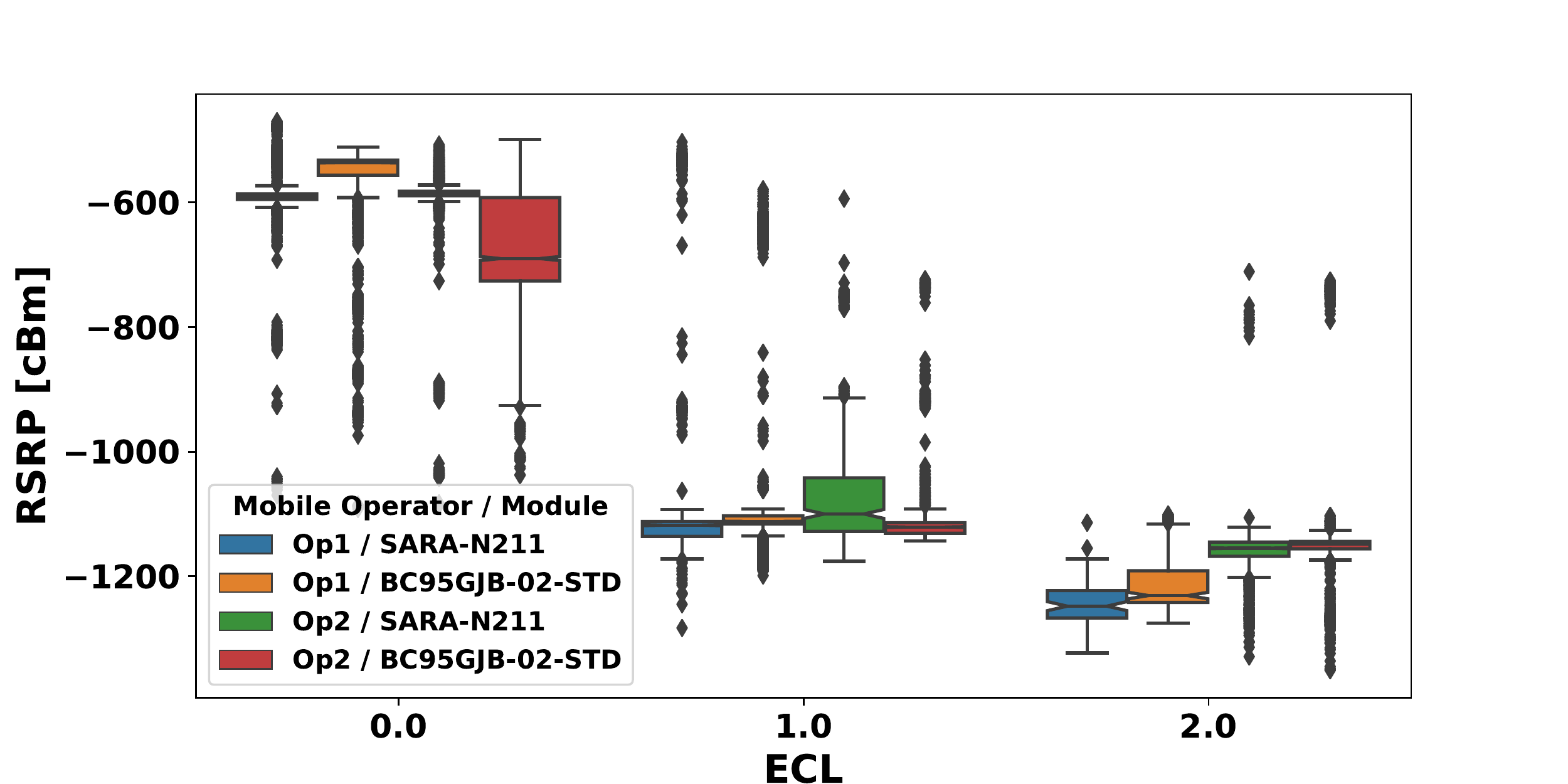}
\caption{Distributions of \gls{rsrp} based on ECL value.}
\label{fig:eclVrsrp} 
\vspace{-2em}
\end{figure}

Fig.~\ref{fig:eclVrsrp} presents the \gls{rsrp} values observed for every
\gls{ecl} level.
From this Figure, we can empirically estimate the \gls{rsrp} thresholds per
operator.
\telia is switching faster to higher \gls{ecl}.
Even though the difference between the thresholds among the operators is small,
as we will present in the next sections, it has a big effect on all the
\gls{kpis}.
Energy consumption and other \gls{kpis} increase marginally between \eclZero and
\eclOne, but deteriorate sharply between between \eclOne and \eclTwo, due to the
huge number of repetitions and use of maximum \gls{txpower}. Some of the metrics
that are affected are: device lifetime, throughput, RTT and packet loss. Using a higher ECL when not necessary, has a
big impact on battery lifetime, without affecting robustness. As we will discuss
in the next chapters, \telia performs poorly in locations with bad coverage.
This is due to its more aggressive \eclTwo threshold.

\begin{figure*}
\centering \includegraphics[trim=0 0 0 40, clip, width=\textwidth]{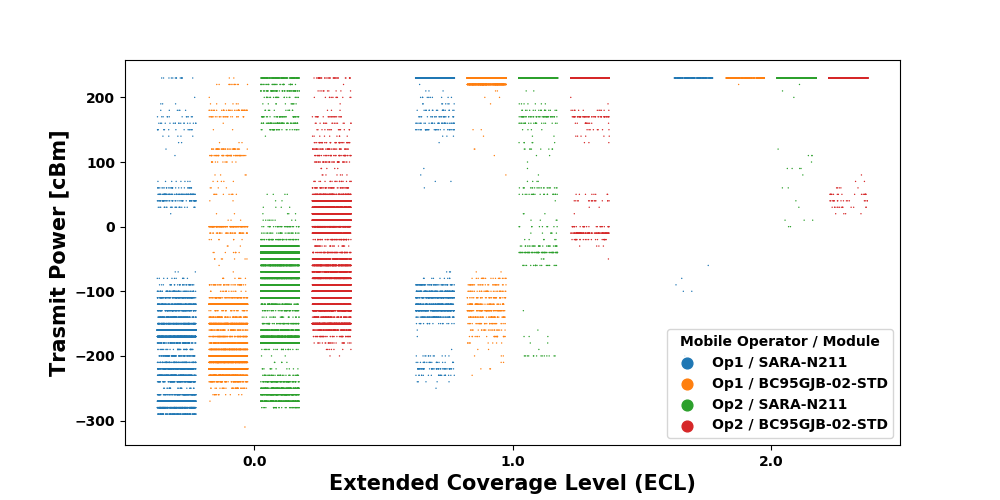}
\caption{Relationship between \gls{ecl} and \gls{txpower}.}
\label{fig:TxPowerVECL}
\vspace{-1em}
\end{figure*}


\begin{figure}
  \centering \includegraphics[trim=0 0 0 40, clip, width=1.1\linewidth]{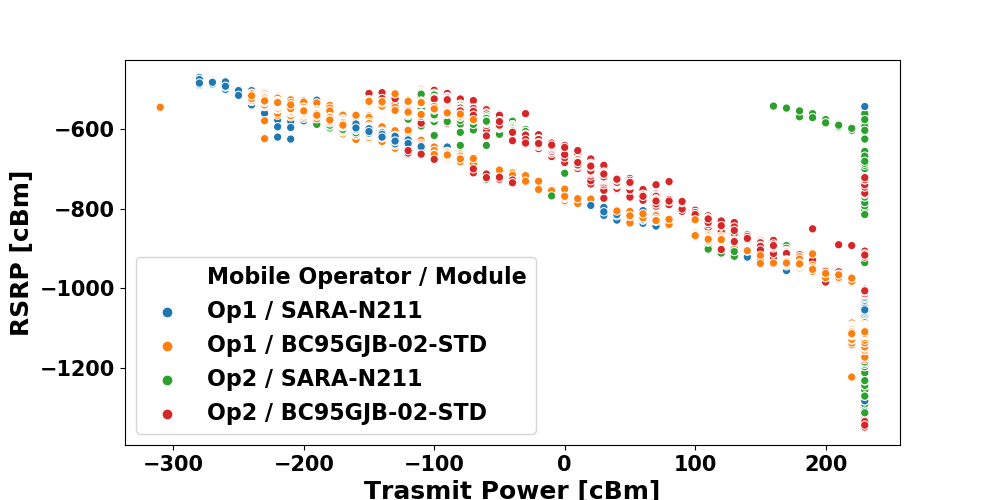}
  \caption{Relationship between \gls{rsrp} and \gls{txpower}.}
  \label{fig:TxVrsrp} 
  \vspace{-1em}
\end{figure}

Finally, we study how \gls{txpower} is connected to \gls{ecl} and \gls{rsrp} in
Fig.~\ref{fig:TxPowerVECL} and~\ref{fig:TxVrsrp}, respectively. In
Fig.~\ref{fig:TxPowerVECL}, we observe the range of possible \gls{txpower} per
\gls{ecl}. Empirical measurements show the minimum \gls{txpower} of a normal LTE
device to be -22 dBm~\cite{joshi2017output}, in contrast the NB-IoT modules may
transmit with as low as -290~cBm ($\approx$ -29~dBm) and the transmit values
have a granularity of 10 cBm. NB-IoT utilizes less bandwidth thus, needs less
\gls{txpower} to reach similar \gls{snr} values to LTE. \eclZero uses the full
range of values and rarely the maximum value of Cat NB1: 230 cBm.
\eclOne uses the maximum value for 79.3\% of the samples.
This is due to the \gls{rach} algorithm discussed above: if the initial value is
\eclZero and the \gls{rach} procedure fails, the \gls{ue} will attempt again with
\eclOne and maximum \gls{txpower}.
As expected, \eclTwo uses
maximum power in 98.4\% of the samples. Fig.~\ref{fig:TxVrsrp} reveals a
linear relationship between \gls{rsrp} and \gls{txpower} and also shows the more
aggressive \gls{txpower} choices of \telia, since for the same \gls{rsrp} value
it usually uses higher \gls{txpower}.
The linear relationship holds for the \gls{rsrp} range typically associated with
\eclZero, between -1000 and -500 cBm. 
Worse \gls{rsrp} values, mostly related with higher ECLs, use almost exclusively
maximum power.

\noindent{\bf Takeaways.} We conclude that of the available metadata metrics,
the most useful are \gls{ecl} and \gls{rsrp}, which are directly related.
Other metrics are either weakly correlated with performance or do not involve
enough variability to be useful.
Operators should carefully choose the mapping between \gls{ecl} and
\gls{rsrp}.

%% file: connected.tex
\section{Energy consumption in Connected state}
\label{sec:energy_consumption_connect}


We now turn to examine whether the actual energy consumption by \gls{nbiot}
\glspl{ue}, while in Connected state, in the real world conforms with the
standard behavior outlined in Sec.~\ref{sec:nbiot_primer}.


\begin{figure}[t]
  \centering
  \includegraphics[trim=0 0 0 25, clip, width=1.1\linewidth]{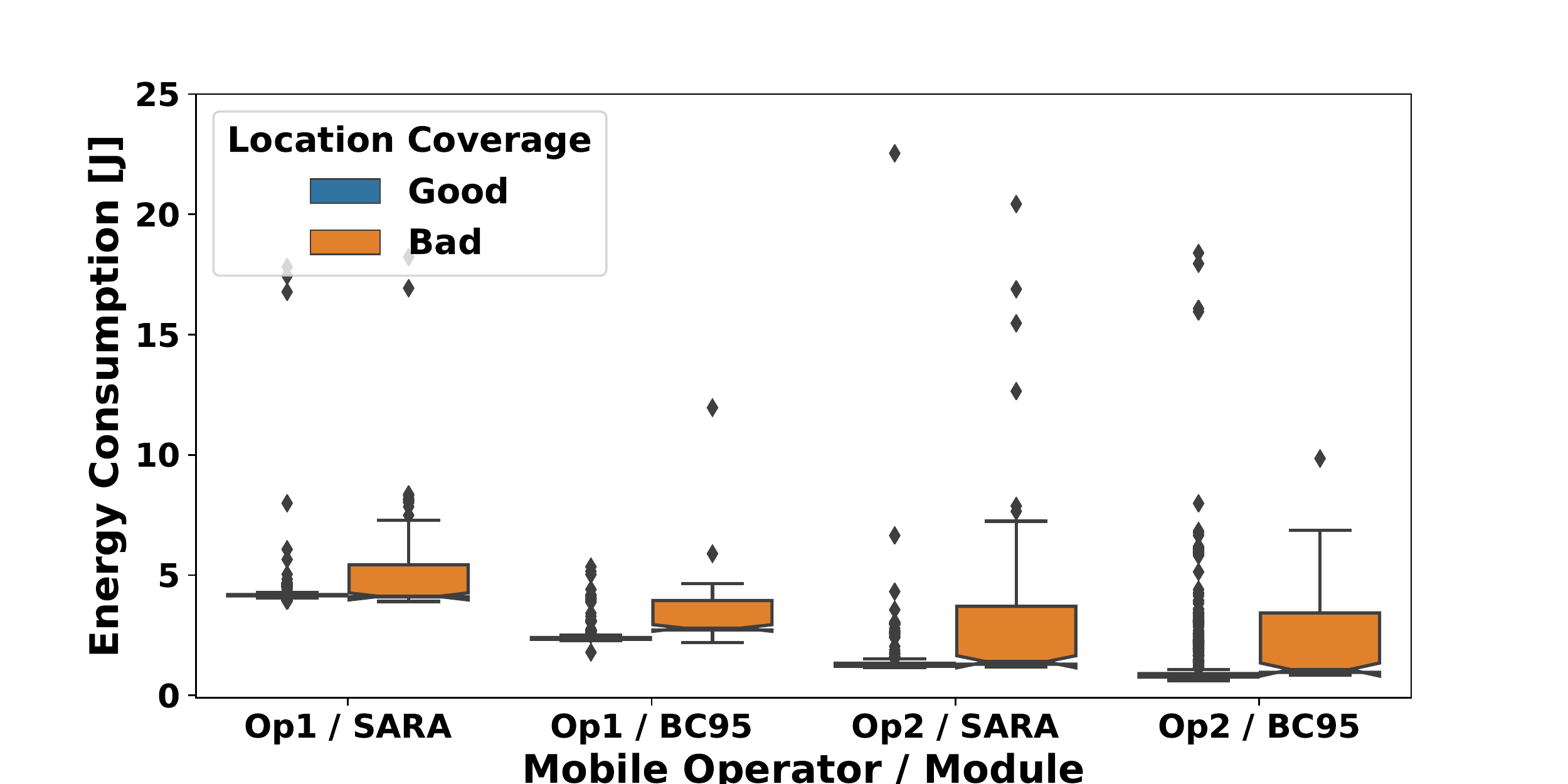}
  \caption{Energy consumption distribution of Connected state, grouped
  by coverage conditions.
  All packet sizes, no \gls{rai}.}
  \label{fig:boxplot_energy_CONNECTED_loc_cov}
\end{figure}

\subsection{Connected state with default settings}
Fig.~\ref{fig:boxplot_energy_CONNECTED_loc_cov} shows the distribution of energy consumption
for the first experiment scenario with no \gls{rai} (\ie \gls{ul} and \gls{dl}
activity with default timers, see Sec.~\ref{sec:exp}) for the different
combinations of operator and module in the Connected state.
In this scenario, we send a UDP packet, which is echoed back by the server.
As we will show in the sequel, packet size has minimal impact in this
configuration, so we include in the default settings analysis all the packet
sizes.
We split the dataset into two groups, depending on the coverage conditions at
the location of the measurements, as discussed in Sec.~\ref{sec:exp_set}.

\begin{figure}
  \captionsetup[subfigure]{justification=centering}
    \begin{subfigure}[t]{\linewidth}
      \includegraphics[trim=0 0 0 40, clip,width=1.1\linewidth]{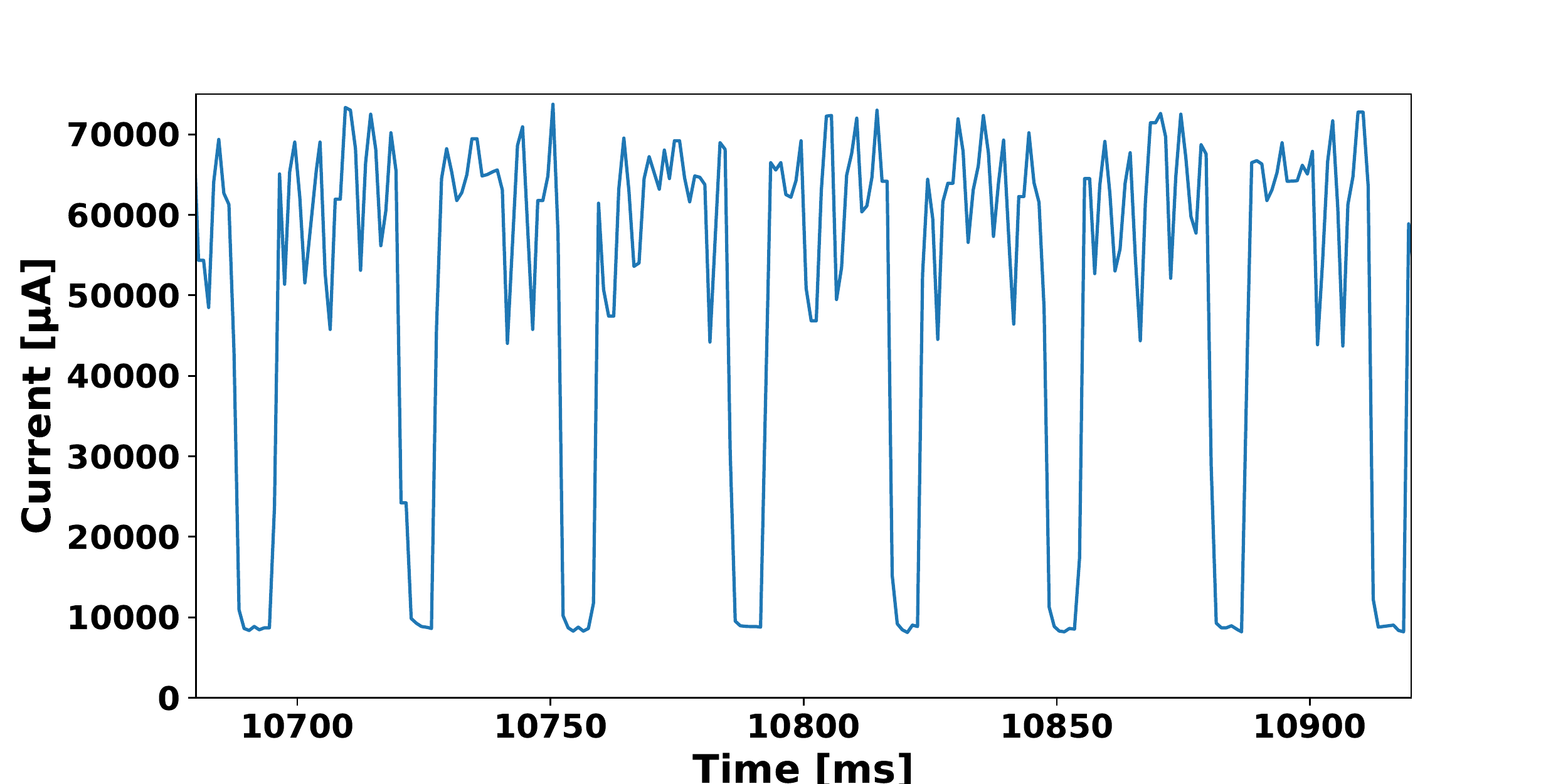}
  \caption{\telenor constantly monitors paging during the inactivity timer.}
  \label{fig:telenor_active_timer}
    \end{subfigure}

    \begin{subfigure}[t]{\linewidth}
      \includegraphics[trim=0 0 0 40, clip,width=1.1\linewidth]{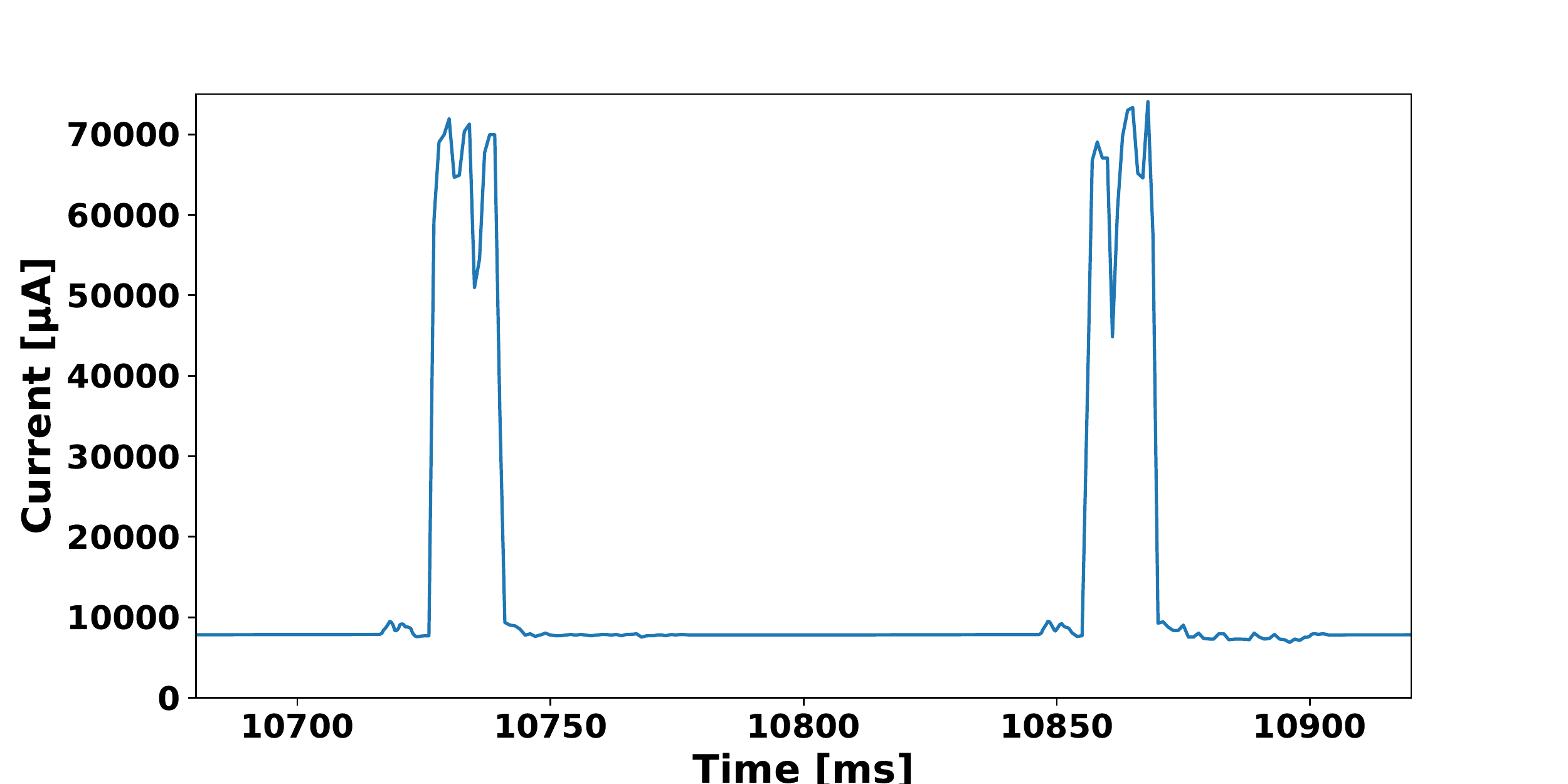}
      \caption{\telia performs cDRX during the inactivity timer.}
      \label{fig:telia_active_timer}
    \end{subfigure}
    \caption{Current draw during the inactivity timer period.}
    \label{fig:activeTimers}
\end{figure}

\noindent{\bf Good coverage.} We record a clear difference between both
operators and modules.
\telenor's energy consumption is 3x or more \telia's.
Also, $\ublox$ consumes more power than $\quectel$, the difference depends on
the operator though!
Table~\ref{tb:uplink_inactivity_good} presents the median energy consumption
for all operator module combinations.
Digging deeper into our data, we find that the difference between the operators
stems from the fact that \telenor does not enforce any quiet period while paging
during the inactivity timer period, like \telia.
Instead, \telenor is mostly in a high energy paging state.
Fig.~\ref{fig:telenor_active_timer} and~\ref{fig:telia_active_timer},
illustrate the behavior of \telenor and \telia respectively, during the
inactivity timer.

\begin{table}[h]
  \caption{Median energy consumption of Connected state under good coverage.
        Includes samples of all packet sizes.}
    \label{tb:uplink_inactivity_good}
    \centering
    \small
    \begin{tabular}{llr}
        \toprule
          Module                & Operator &  Energy [J] \\
        \midrule
\multirow{2}{*}{Quectel - BC95} & Op1      &  2.39       \\
                                & Op2      &  0.82       \\
\cline{1-3}
\multirow{2}{*}{SARA - N211}    & Op1      &  4.17       \\
                                & Op2      &  1.27       \\
\bottomrule
        \end{tabular}
        
\end{table}

\begin{table}[h]
  \caption{Median energy consumption of Connected state under poor coverage.
        Includes samples of all packet sizes.}
    \label{tb:uplink_inactivity_bad}
    \centering
    \begin{tabular}{lllr}
        \toprule
    Module & Operator & ECL &  Energy [J] \\
      \midrule
      \multirow{6}{*}{Quectel - BC95} & \multirow{3}{*}{Op1} & 0 &               2.71  \\
                  &     & 1 &               2.80\\
                  &     & 2 &               4.04\\
      \cline{2-4}
                  & \multirow{3}{*}{Op2} & 0 &               0.88 \\
                  &     & 1 &               1.03 \\
                  &     & 2 &               3.44 \\
      \cline{1-4}
      \cline{2-4}
      \multirow{6}{*}{SARA - N211} & \multirow{3}{*}{Op1} & 0 &               4.15 \\
                  &     & 1 &               4.10  \\
                  &     & 2 &               5.50  \\
      \cline{2-4}
                  & \multirow{3}{*}{Op2} & 0 &               1.28 \\
                  &     & 1 &               1.40  \\
                  &     & 2 &               3.77  \\
      \bottomrule
        \end{tabular}
        
\end{table}

\begin{figure}
  \centering
  \includegraphics[trim=0 0 0 40, clip,width=1.1\linewidth]{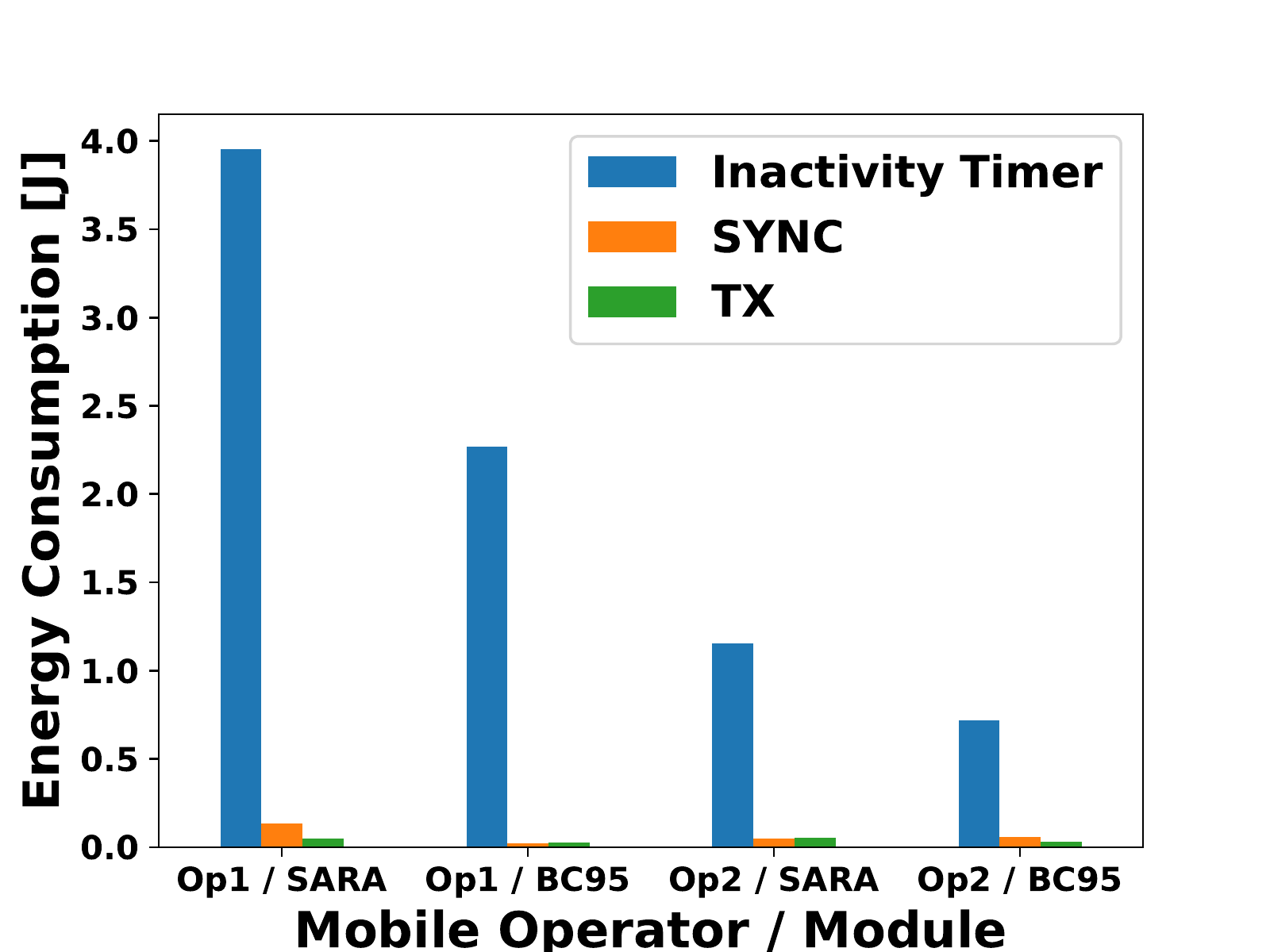}
  \caption{Break down of energy consumption per Connected state substate for
  locations with good coverage.
  We used the median value of each substate.
  All packet sizes.}
  \label{fig:connected_substates_energy}
   
\end{figure}

\begin{figure}
  \centering
  \includegraphics[width=1.1\linewidth]{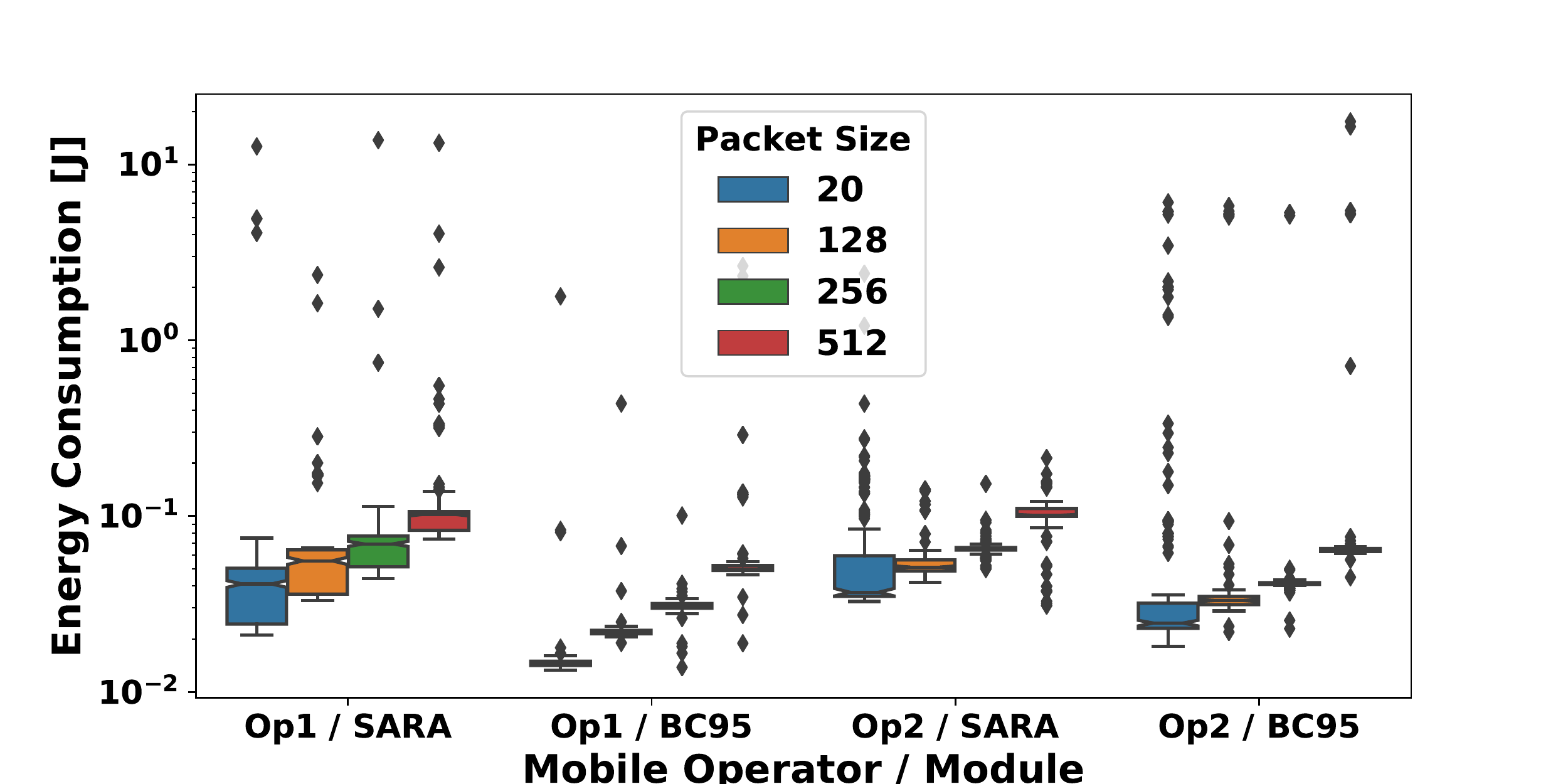}
  \caption{Distributions of the consumed energy of the TX substate, grouped
  by the packet size of the transmission, for locations with good coverage.
  Y axis is in logarithmic scale.
  }
  \label{fig:Tx_good_pSize_energy}
\end{figure}

Fig.~\ref{fig:connected_substates_energy} shows the median values of the
consumed energy for every substate of the Connected state.
Recall that the Connected state comprises three substates: synchronization with
the network (sync), data plane transmission and reception (TX) and inactivity
timer.
In the inactivity timer, the \gls{ue} performs paging and, ideally,
enforces \gls{cdrx}.
The inactivity timer substate dominates the energy consumption. Thus, it is
critical to consider whether it is necessary, and if so, \gls{cdrx} should be
used. This also hints at that the size of the transmitted packet becomes
irrelevant, since the increase in energy consumption for the extra bytes is
minuscule compared to the total energy consumption of the Connected state.
Fig.~\ref{fig:Tx_good_pSize_energy} illustrates the energy consumption for
different packet sizes. The cost increases sub-linearly with packet size -- the
bigger a packet is, the less energy is spent per byte. Increasing the packet
size from 20 to 512 bytes results in an increase in energy consumption by a few
tens of mJ, negligible when compared to the energy consumed in the inactivity
timer substate, which is in the order of Joules.

\smallskip
\noindent{\bf Poor coverage.} Fig.~\ref{fig:boxplot_energy_CONNECTED_loc_cov}
shows no clear differences in the median power consumption between locations
with good and poor coverage.
However, the latter are characterized by stronger variability with the
inter-quartile difference several times the median.
The difference in coverage results in picking different ECL levels.
A higher ECL means extra repetitions when sending data, to increase the
likelihood of successful delivery, which translates into a higher energy
consumption.
In brief, there is a slight increase in energy consumption between \eclZero and
\eclOne and a big increase between \eclOne and \eclTwo.
The devices at good signal locations are using almost always \eclZero and in about
1\% of the cases \eclOne.
This explains the compactness of the energy consumption distribution.
At bad signal locations the devices are on \eclTwo for about 30\% of the samples,
with the rest of the samples split between \eclZero and \eclOne.
These proportions vary according to the device and operator combinations.
This 70/30 split explains why the median energy consumption in poor coverage is
close to that in good coverage.
Table~\ref{tb:uplink_inactivity_bad} presents the median energy consumption
for all operator module combinations split further according
to the ECL level.
The values in the table confirm that the difference between poor and good
coverage is chiefly evident for \eclTwo.

\subsection{ Connected state with \gls{rai}.}
\label{sec:connected-rai}
Now we move to discuss the energy consumption when the \gls{rai} flag is set.

\begin{figure}[t]
  \centering
  \includegraphics[width=1.1\linewidth]{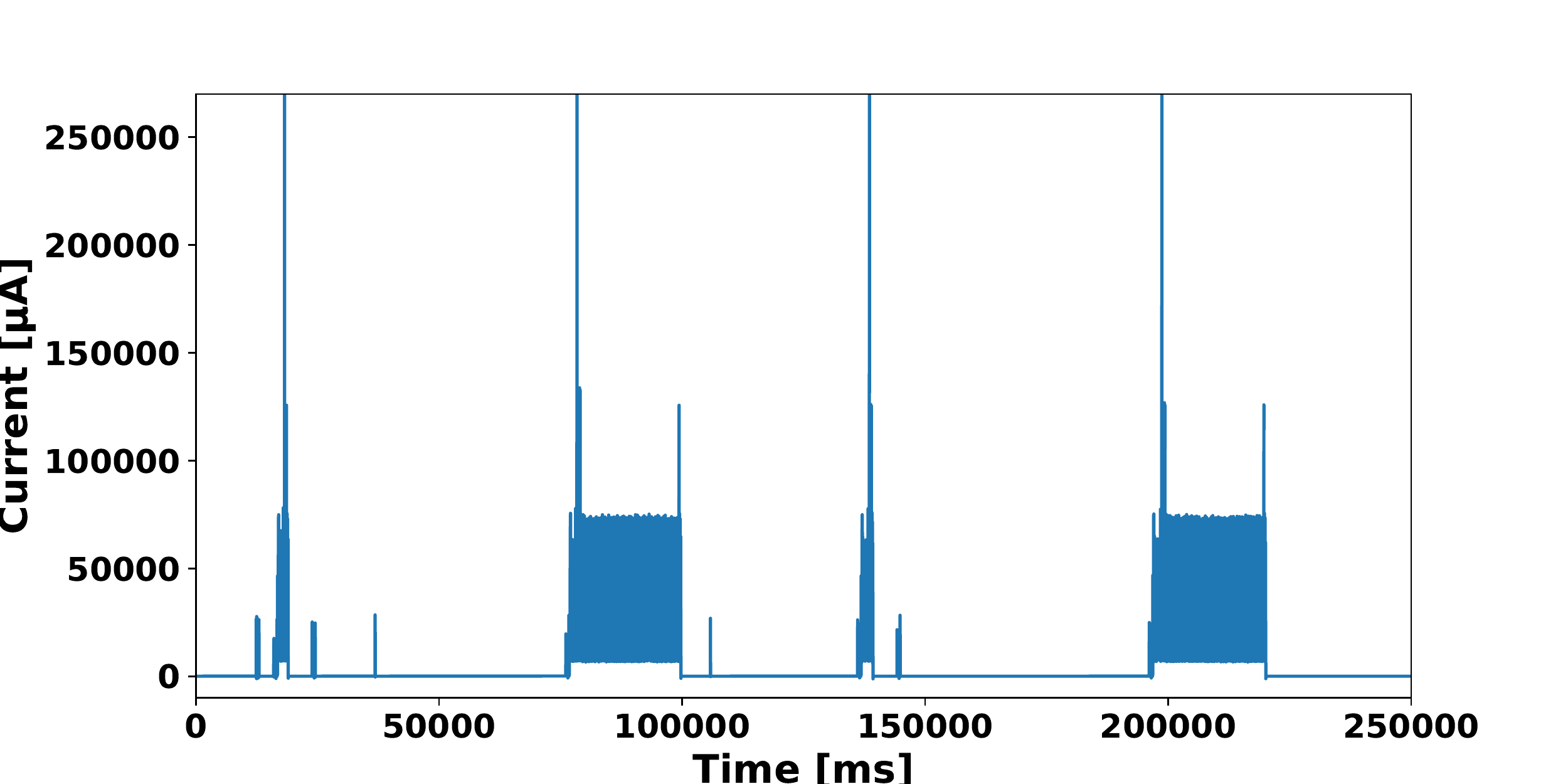}
  \caption{Current draw of \telia over SARA-N211 when using the RAI-200
  flag in early 2019. 20 bytes packets. Good coverage.}
  \label{fig:telia_rai_200_bug}
\end{figure}

\begin{figure}[t]
  \centering
  \includegraphics[width=1.1\linewidth]{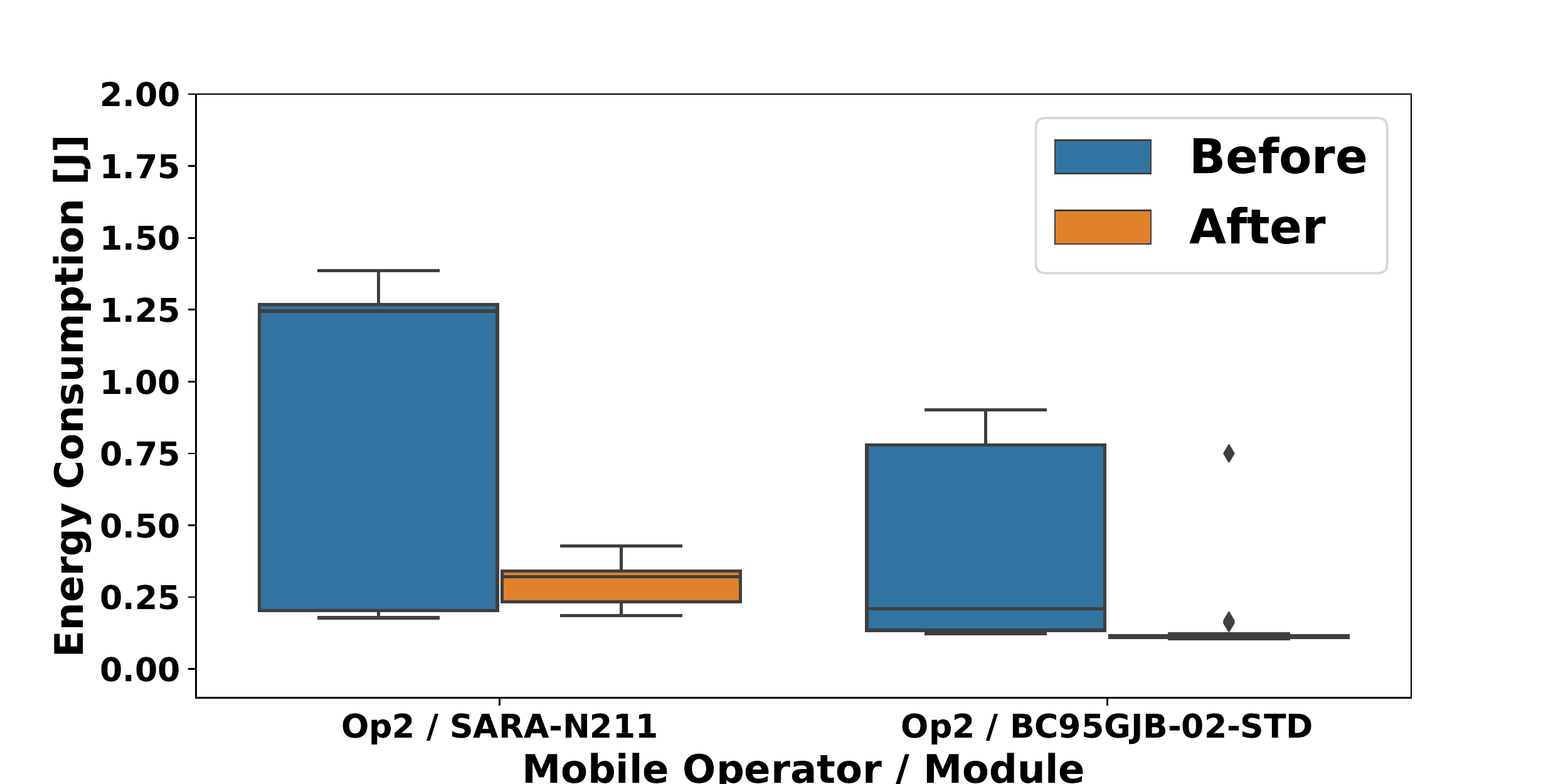}
  \caption{Energy consumption of \telia when sending 20 bytes using \gls{rai}-200
  in a location with good signal before
  and after our feedback.}
  \label{fig:telia_rai_200_feedback}
\end{figure}

\smallskip
{\noindent{\em Is the \gls{rai} flag respected?}}  We observe that both operators may
ignore the flag and proceed to perform an inactivity timer procedure.
For \telenor, this is a rare occurrence, it just happened for 3 packets in our
dataset.
A plausible cause could be corrupt signaling packets.
For \telia, however, \gls{rai}-200 flag was ignored, for 50\% of all packets of
all measurements performed before April 2019, regardless of the module. More
specifically, one every two packets would consistently utilize the inactivity
timer, after transmission, instead of dropping to Idle state.
Fig.~\ref{fig:telia_rai_200_bug} showcases this behavior. The short spikes are
transmissions where the flag was respected, while the periods with intense
activity (\eg the one starting around $t=200000$) are instances where the flag
was ignored. Following the discovery of this anomaly, we informed \telia of it.
The operator then corrected the misconfiguration that caused it.
Fig.~\ref{fig:telia_rai_200_feedback} shows the energy consumed to transmit 20
bytes, while setting \gls{rai}-200, before and after our feedback to the
operator. In the ``before'' case,
the distributions are broader exhibiting values similar to those measured when
the \gls{rai} is not in use (see Fig.~\ref{fig:boxplot_energy_CONNECTED_loc_cov}).
fixing this bug has led to a reduction in the median value by 80\%. The gains
are even higher for larger transmissions and/or challenging signal conditions.
In the rest of the Section, we only present measurements collected after the
correction.

\begin{figure*}[]
  \centering
  \includegraphics[width=0.85\paperwidth]{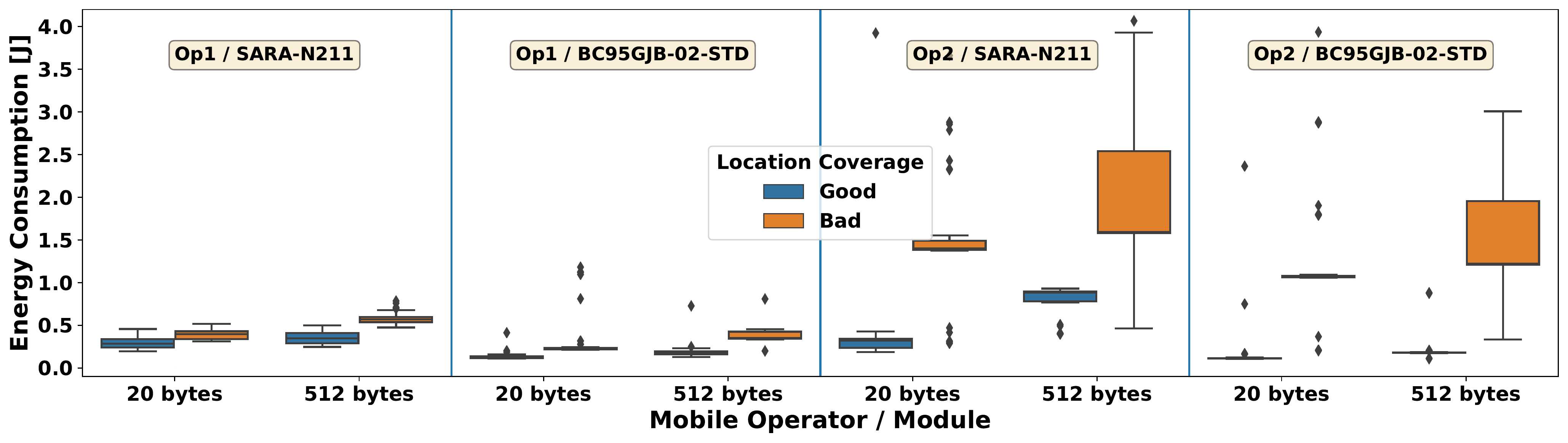}
  \caption{ Distribution of energy consumption for \gls{rai}-200, grouped by
  coverage conditions and packet size.}
  \label{fig:boxplot_energy_rai_200_loc_cov}
\end{figure*}

\begin{figure*}[]
  \centering
  \includegraphics[width=0.85\paperwidth]{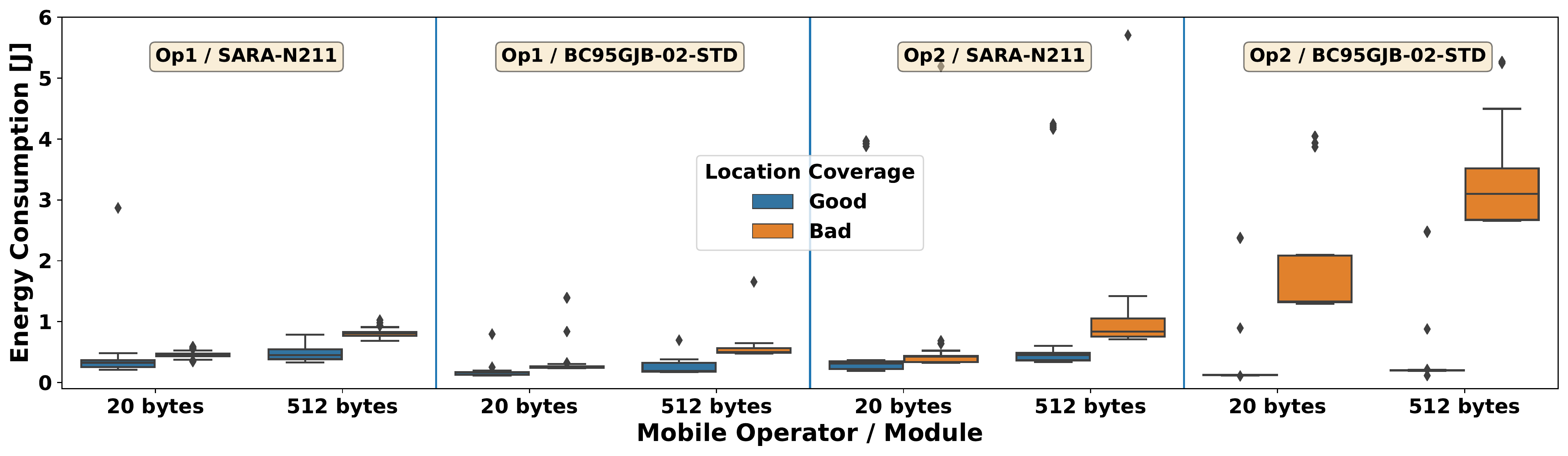}
  \caption{ Distribution of energy consumption for \gls{rai}-400, grouped by
  coverage conditions and packet size.}
  \label{fig:boxplot_energy_rai_400_loc_cov}
\end{figure*}

\smallskip
{\noindent{\em Energy consumption.}} Fig.~\ref{fig:boxplot_energy_rai_200_loc_cov}
and~\ref{fig:boxplot_energy_rai_400_loc_cov} present the energy consumption for
several combinations of packet sizes and coverage locations while setting the
flags \gls{rai}-200 and \gls{rai}-400, respectively.
Using \gls{rai} leads to great savings in energy.
Table~\ref{tb:uplink_rai_good} presents the median energy consumption, for all
operator, module and \gls{rai} combinations.
The median energy consumption without \gls{rai} can be 20x and 15x that with \gls{rai}
for \telenor and \telia, respectively.
Furthermore, the choice of the module influences the energy consumption
greatly.
$\ublox$ consumes about double that of $\quectel$.

Interestingly, using \gls{rai}-400 results, in most cases, in only a slight
extra energy consumption compared to \gls{rai}-200. This may partially be
attributed to the locality of the echo server, offering short round trip times,
and by extension reducing the total duration of the Connected state.
For example, if we consider the best performing pair of operator and module:
\telenor and $\quectel$, under good conditions.
When sending a 20-byte packet, the median Connected state duration (\ie time
between two Idle states), is 3.13 seconds for \gls{rai}-200 and 3.23 seconds for
\gls{rai}-400.
The median duration when transmitting 512 bytes remains unaffected at 3.12
seconds with \gls{rai}-200, but reaches 4.06 seconds with \gls{rai}-400.
Under poor coverage, transmitting 512 bytes requires 3.94 (\gls{rai}-200) or
4.95 seconds (\gls{rai}-400).

\begin{table}[h]
  \caption{Median energy consumption when sending 20 bytes with \gls{rai} under good coverage (Joules).}
  \label{tb:uplink_rai_good}
    \centering
    \small
    \begin{tabular}{llrr}
        \toprule
        module                      & operator & \gls{rai}-200 & \gls{rai}-400 \\
        \midrule

        \multirow{2}{*}{$\quectel$} & \telenor & 0.12          &  0.17         \\
                                    & \telia   & 0.11          &  0.12         \\
        \cline{1-4}
        \multirow{2}{*}{$\ublox$}   & \telenor & 0.27          &  0.31         \\
                                    & \telia   & 0.31          &  0.33         \\
        \bottomrule
        \end{tabular}
        

\end{table}

With the inactivity timer substate being removed, the impact of the payload size
increases.
We have tested two very different packet sizes: 20 and 512 bytes (see
Fig.~\ref{fig:boxplot_energy_rai_200_loc_cov}
and~\ref{fig:boxplot_energy_rai_400_loc_cov}).
The larger packets result in a larger energy consumption.
This increase hovers around 60\% and never exceeds 100\%.
Hence, although payload size plays an important role, the choice of the module
has more impact.
For example, if we focus on the the median energy consumption of \telenor over
\gls{rai}-400 (\ie the two left quadrants
of Fig.~\ref{fig:boxplot_energy_rai_400_loc_cov}), keeping the packet size
constant and changing the module from $\quectel$ to $\ublox$
results in an $115\%$ increase in good locations and a $70\%$
increase in bad locations.
Finally, we observe that \telia draws significantly more power, at places with
poor coverage, compared to \telenor.
Digging deeper into this, we find that \telia uses \eclTwo more frequently
than \telenor, as was expected based on \telia's more aggressive \gls{ecl}
thresholds, we detected in Fig.~\ref{fig:eclVrsrp} of
Section~\ref{sec:metadata_quality}.
This results in repeating each transmission several times, causing an up to
tenfold increase of the overall energy cost compared to \telenor under similar
conditions.

\noindent{\bf Takeaways.} The use of \gls{rai} flag leads to significant savings
in energy consumption.
The choice of the \gls{ue} is key to energy consumption, which suggests the need
for a \gls{ue} certification process.
Operators must thoroughly test and confirm that their implementation and
configuration is conform with the expected standard behavior.
We have highlighted a few cases of misconfiguration that translate into
excessive energy consumption.
Interestingly, the measured \gls{nbiot} deployments seem to fare well under poor
coverage conditions except for the extremes.
Payload size becomes important only when the \gls{rai} flag is set
and the network is correctly configured.

%% file: idle.tex
\section{Power consumption in Idle state}
\label{sec:energy_consumption_idle}

The majority of the NB-IoT device's lifetime is spent on Idle state and mostly
on \gls{psm} and the sleep phase of \gls{idrx}, if available.
This section quantifies power consumption in the \gls{psm} and \gls{edrx}
modes.\footnote{\gls{idrx} can be either \gls{drx} or \gls{edrx} (\ie \gls{drx}
with PTW and prolonged sleep periods). Our analysis applies to both, but we use
\gls{edrx} in this Section,
since this is expected to be more popular in NB-IoT.}
Note that these modes do not have a specific time duration, thus we present the
power consumption rather than the energy.

\subsection{PSM}
\label{sec:energy_consumption_idle_psm}
During \gls{psm}, the radio is OFF and the device is in a ``deep sleep'' mode.
Thus, the only parameter affecting power consumption is the module itself (\ie
the combination of the hardware and firmware).
Both modules consume around 10~$\mu$W, with the median values being
10.61~$\mu$W for $\quectel$ and 9.35 $\mu$W for $\ublox$. 
In rare occasions ($< 2\%$ of the \gls{psm} samples in the dataset), the modules
fail to reach the typical \gls{psm} current levels of 2-5~$\mu$A, resulting in
an elevated power consumption that may exceed 30~$\mu$W. Hence, the power
distributions (not shown) are fairly compact, with 98\% of all samples centered
around the median.


\subsection{eDRX}
\label{sec:energy_consumption_idle_edrx}
\gls{edrx} consists of listening and sleep phases (see
Sec.~\ref{sec:operational_phases}).
The \textit{eDRXCycle} parameter determines the overall duration of an
\gls{edrx} cycle, which is the time between the starting points of two
consecutive listening phases.
However, the total duration of the sleep phase is not standardized, because the
listening phase may vary in length due to channel conditions.
To estimate the energy consumption, while on \gls{edrx}, we measure the time
spent listening $t_{eDRX-L}$ as well as the consumed power $P_{eDRX-L}$.
Multiplying these two gives the energy consumed while listening: $E_{eDRX-L}$.
The time spent sleeping equals the total time spent in \gls{edrx} minus
the time spent listening ($t_{eDRX-S} = t_{eDRX-total} - t_{eDRX-L}$).
The power consumed while sleeping ($P_{eDRX-S}$), is the same as range as
\gls{psm}.
Hence, the overall energy consumption in \gls{edrx} is given by:

\begin{equation}
    \label{eq:edrx}
    E_{eDRX} = (E_{eDRX-L} + P_{eDRX-S} \cdot t_{eDRX-S}) \cdot N_{cycles},
\end{equation}

where $N_{cycles}$ is the number of listening-sleep cycles in the eDRX mode,
which can be derived by the configuration.
Next, we examine each of the two phases.

\subsubsection{Listening}



The duration of the listening phase depends chiefly on coverage conditions,
with listening phases in bad coverage lasting significantly longer.
More specifically, it is affected by the ECL, which is in turn determines
the number of control channel repetitions the UE should listen.
Table~\ref{tb:idrx_on_model} presents the median values of $t_{eDRX-L}$ and
$E_{eDRX-L}$ for different operator, module and coverage combinations.

\begin{figure}[t]
     \centering
     \includegraphics[width=1.1\linewidth]{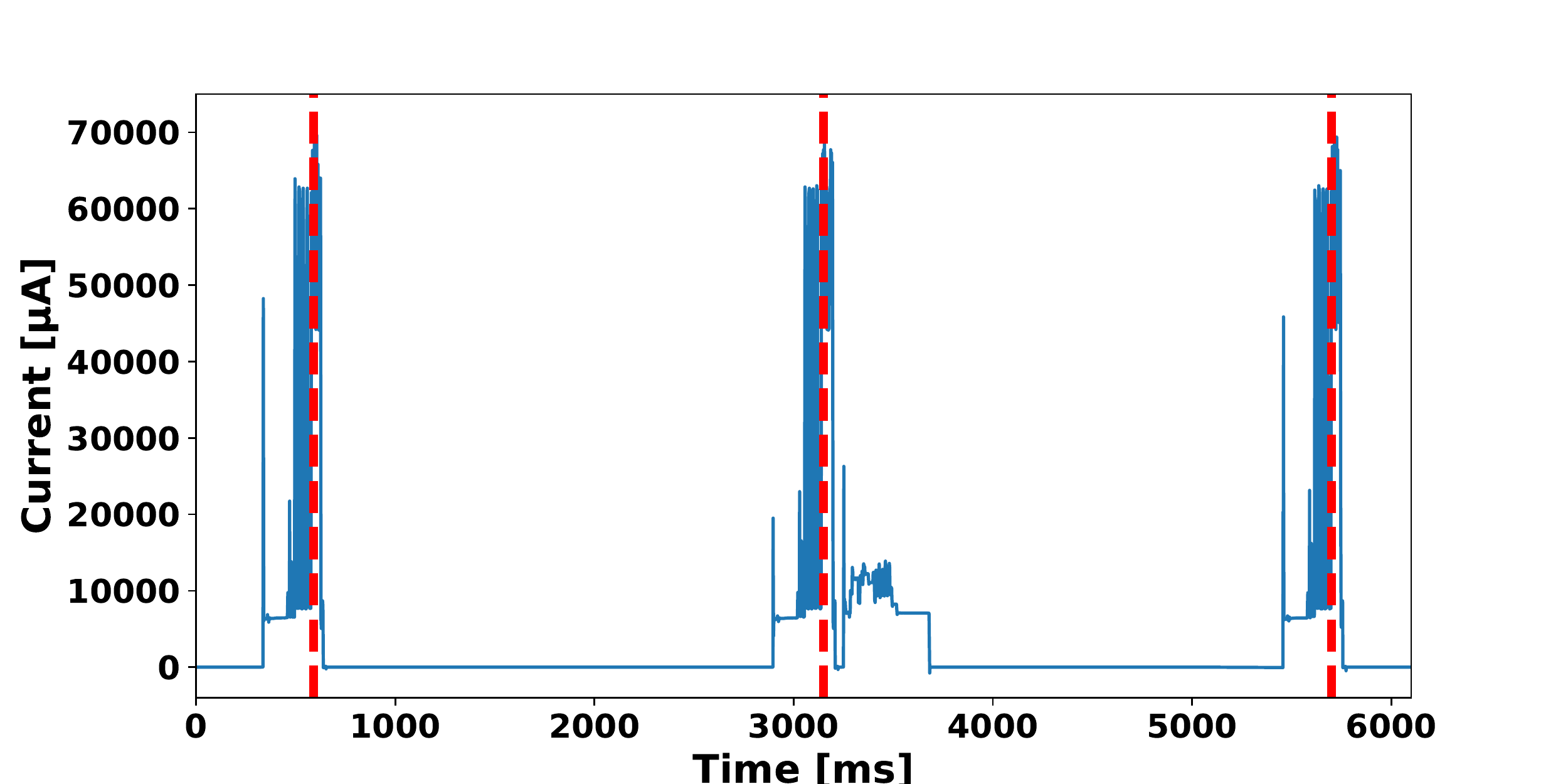}
     \caption{Current draw of the buggy \gls{edrx} listening phase implementation.
     The listening phases should end at the dashed vertical lines.}
     \label{fig:telenor_idrx_ON_bug}
   \end{figure}

Listening starts with a low power synchronization period and ends with a more
power demanding period of listening to paging occasions (PO).
There is a bug observed in both devices, where they might remain at an
elevated power level after the PO period ends, shown in
Fig.~\ref{fig:telenor_idrx_ON_bug}, increasing the phase's duration and
energy.
The proper ending points of the PO periods are marked with red lines in the
Figure.
This bug appeared mostly when using \ublox over \telenor and in later
measurements appears much less frequently.
Table~\ref{tb:idrx_on_model} uses only recent measurements, where the bug is
rarely observed (in the parenthesis we present the values while the bug was
still frequent).

\begin{table}[]
     \caption{Median values of energy consumption and duration of
        \gls{edrx} listening phase.}
    \label{tb:idrx_on_model}
    \centering
    \begin{tabular}{lllrr}
        \toprule
            Coverage & Module & Operator &  Energy [mJ] & Duration [ms]\\
            \midrule
\multirow{4}{*}{Bad} & \multirow{2}{*}{Quectel - BC95} & Op1 &   21.4 &  470.2 \\
     &             & Op2 &   24.6 &  476.7 \\
\cline{2-5}
     & \multirow{2}{*}{SARA - N211} & Op1 &   33.7 &  536.5 \\
     &             & Op2 &   39.1 &  552.2 \\
\cline{1-5}
\cline{2-5}
\multirow{4}{*}{Good} & \multirow{2}{*}{Quectel - BC95} & Op1 &    6.4 &  215.0 \\
     &             & Op2 &    6.3 &  215.2 \\
\cline{2-5}
     & \multirow{2}{*}{SARA - N211} & Op1 &   10.3 (20.0) & 224.5 (300.0) \\
     &             & Op2 &   10.1 &  222.8 \\
\bottomrule
        \end{tabular}
        
\end{table}

Under good conditions we do not observe any difference between the operators for
the same module. The device though has a major effect in energy consumption,
with \ublox consuming a median 10 mJ and \quectel a median 6 mJ. Under bad
conditions, the power consumption mostly depends on the ECL.
\telia has a
tendency to switch to \eclTwo faster than \telenor, as conditions get worse, and
this is reflected in the energy consumption. As in good conditions, the most
important factor in the energy consumption is the module, with \quectel showing
better efficiency. Table~\ref{tb:idrx_on_model} also shows that listening
time duration evidently increases under poor coverage.

\subsubsection{Sleeping}
As with \gls{psm}, the deciding factor of the energy consumption in the sleeping
phase is the module. The median of $P_{eDRX-S}$ is 10.01 $\mu$W and 10.36
$\mu$W for $\ublox$ and $\quectel$, respectively.

\noindent{\bf Takeaways.} In deep sleep, energy consumption is only affected by
the choice of the module. Energy consumption while listening is determined by
coverage and the choice of the module under good conditions. Under bad
conditions, operator choice becomes important as well.

%% file: rttPacketLoss.tex
\section{Network Performance: \gls{rtt}, Throughput and Packet Loss}
\label{sec:rttPacketLoss}

\begin{table*}

    \caption{Summary of \gls{kpis}.
    Packet Loss is calculated among the whole dataset.
    Throughput and \gls{rtt} are calculated for 20 bytes packets.}
    \label{tab:performanceSummaryTable}
    \begin{adjustbox}{max width=\textwidth}
    \begin{tabular}{|l|l|r|r|r|r|r|r|r|}
        \hline
        \pbox{5cm}{Mobile Operator / Module} & \pbox{5cm}{Location\\Coverage} & \pbox{5cm}{Packet Loss:\\Both Directions [\%]} & \pbox{5cm}{Packet Loss:\\Uplink [\%]} & \pbox{5cm}{Packet Loss:\\Downlink [\%]} & \pbox{5cm}{\gls{rtt} Median\\20 bytes [S]} & \pbox{5cm}{Throughput Median\\20 bytes [bps]} & \pbox{5cm}{Energy Consumption default\\Median all packet sizes [J]} & \pbox{5cm}{Energy Consumption RAI-400\\Median 20 bytes [J]} \\
        \hline
            Op1 / BC95GJB-02-STD &               Bad &                            0.621 &                   0.573 &                     0.048 &                   3.333 &                          386.708 &                                              2.801 &                                          0.251 \\
            Op1 / BC95GJB-02-STD &              Good &                            0.468 &                   0.401 &                     0.067 &                   2.894 &                         1574.426 &                                              2.388 &                                          0.162 \\
                 Op1 / SARA-N211 &               Bad &                            0.427 &                   0.047 &                     0.379 &                   4.045 &                          385.890 &                                              4.114 &                                          0.447 \\
                 Op1 / SARA-N211 &              Good &                            0.124 &                   0.093 &                     0.031 &                   3.102 &                         1235.531 &                                              4.174 &                                          0.326 \\
            Op2 / BC95GJB-02-STD &               Bad &                            1.130 &                   0.963 &                     0.167 &                   5.391 &                          203.304 &                                              1.048 &                                          1.331 \\
            Op2 / BC95GJB-02-STD &              Good &                            0.947 &                   0.726 &                     0.221 &                   2.753 &                          400.751 &                                              0.819 &                                          0.121 \\
                 Op2 / SARA-N211 &               Bad &                            0.855 &                   0.617 &                     0.237 &                   6.220 &                          203.110 &                                              1.396 &                                          0.427 \\
                 Op2 / SARA-N211 &              Good &                            0.772 &                   0.386 &                     0.386 &                   2.478 &                          399.750 &                                              1.270 &                                          0.311 \\
        \hline
        \end{tabular}
    \end{adjustbox}
    \vspace{-2em}
\end{table*}

Finally, we examine the network \gls{kpis}: packet loss, \gls{rtt} and
throughput. Table~\ref{tab:performanceSummaryTable}, presents a summary of these
metrics, as well as some of the metrics discussed in the previous sections,
allowing for a complete overview of the performance.

\noindent{\bf Packet Loss.}
In our experiments we transmit a single UDP packet to a well provisioned
server and, if applicable, echo it back to the device. We embed each
packet with a unique ID. If the packet never reaches the server we assume it was
lost in the \gls{ul} direction. In the experiments where the UE is expecting a
response, if a packet reaches the server but the corresponding reply is never
received by the UE, we assume a loss in the \gls{dl} direction.
LTE UEs (\eg smartphones) experience almost null packet loss when they are
immobile / stationary and connected to uncongested LTE networks
(~\cite{techreport:techreptype}, Fig. 3.1 and Tab. 3.1).
In contrast, we observe that packet loss rates in commercial NB-IoT
deployments are between 0.5\% and 1\%. The majority of the losses happen in the
\gls{ul}, and worsening signal conditions cause a slight increase, as expected.
Surprisingly, the more aggressive use of robust ECL levels by \telia, does not
translate into better packet delivery, compared to \telenor. This might indicate
that the losses are not happening in the Radio Access Network. If guaranteed
delivery is important, the use of a higher layer protocol such as MQTT or CoAP
is needed.

\begin{figure}
    \includegraphics[width=1.1\linewidth]{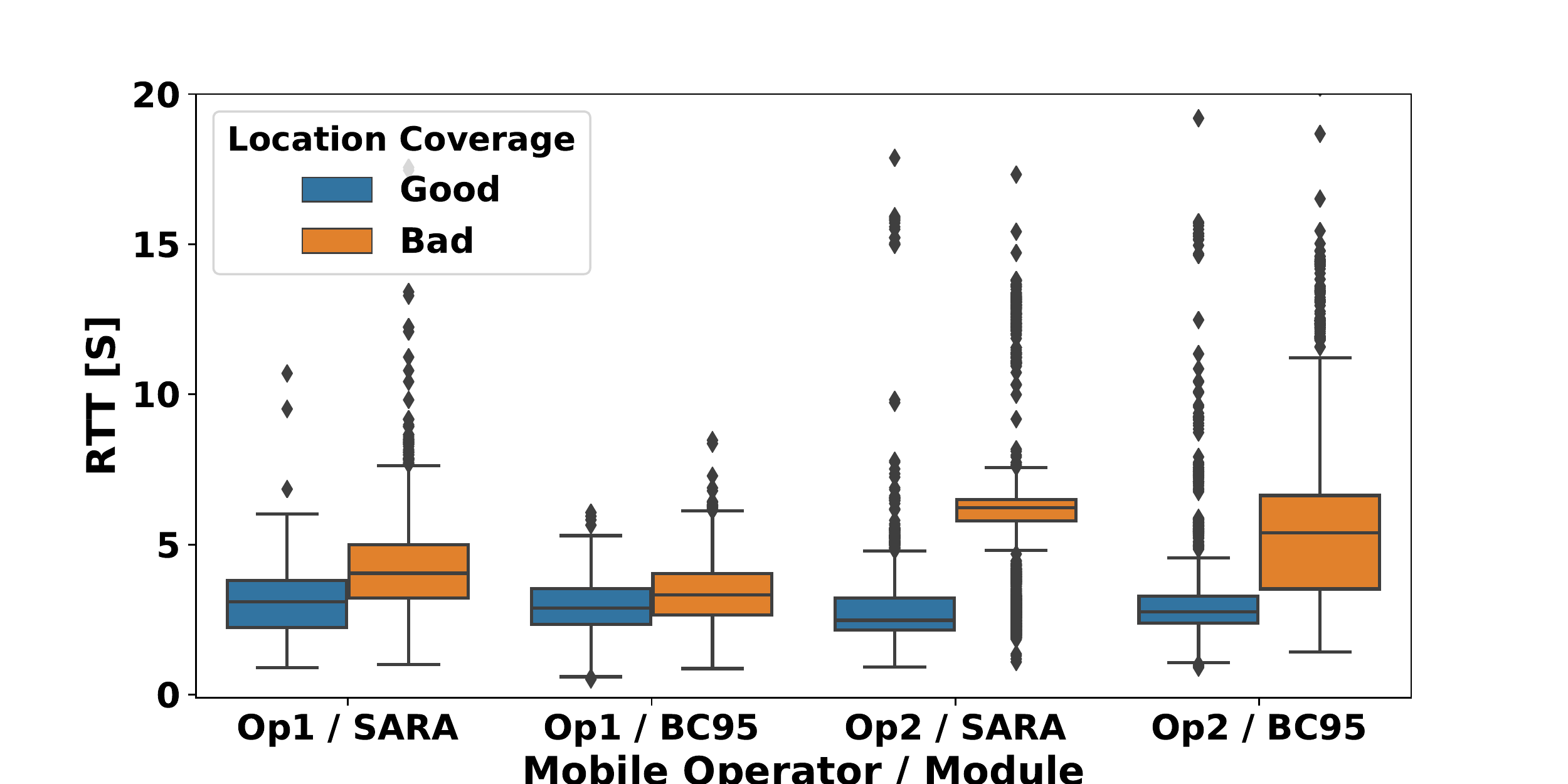}
    \caption{\gls{rtt} per location coverage when sending a 20-byte packet.
    All packet sizes.}
    \label{fig:RTTVlocationCoverage} 
    \vspace{-2em}
\end{figure}

\begin{figure}
    \includegraphics[width=1.1\linewidth]{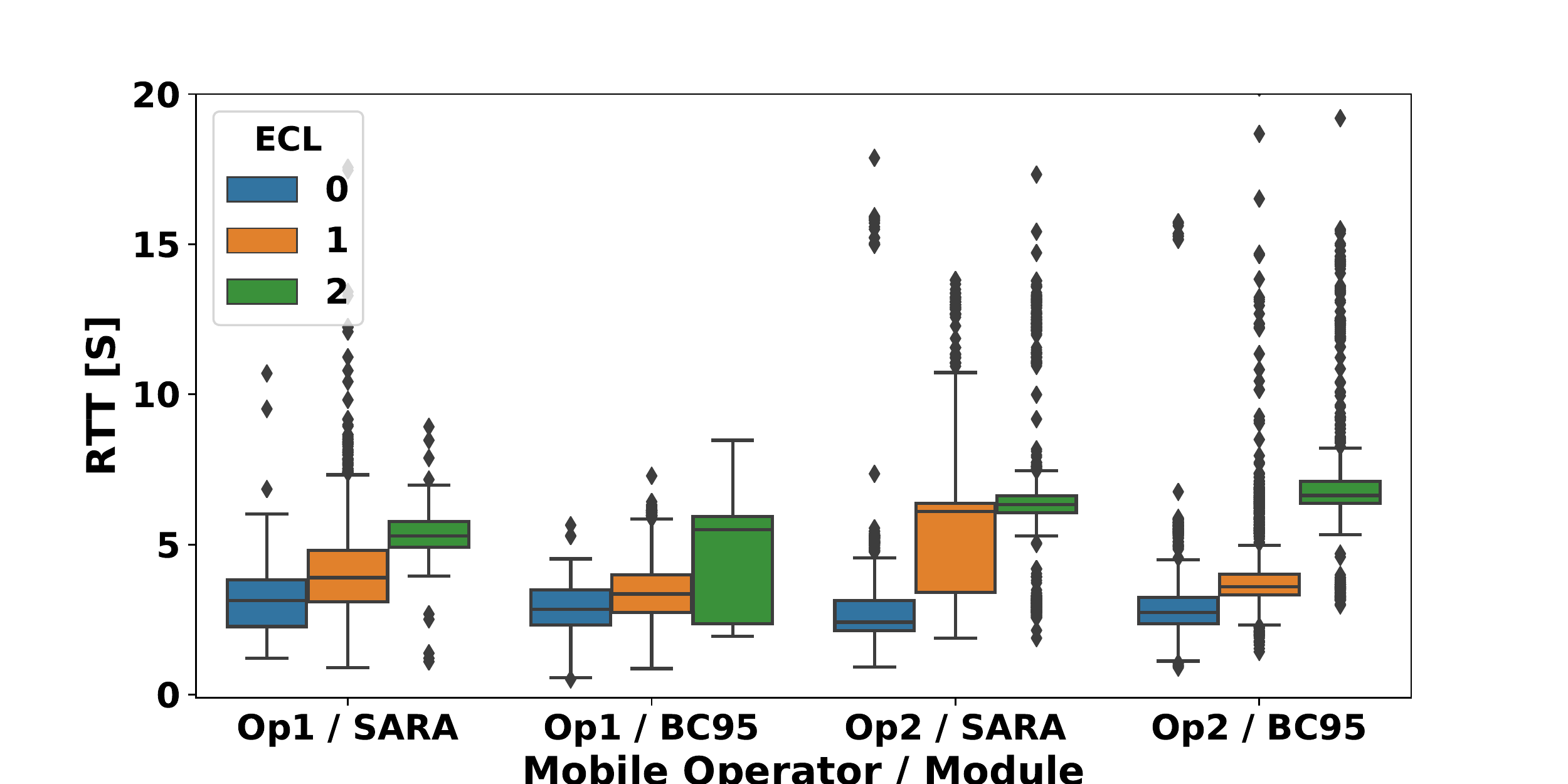}
    \caption{\gls{rtt} per ECL level when sending a 20-byte packet.}
    \label{fig:rttVecl} 
    \vspace{-2em}
\end{figure}

\begin{figure}
    \includegraphics[width=1.1\linewidth]{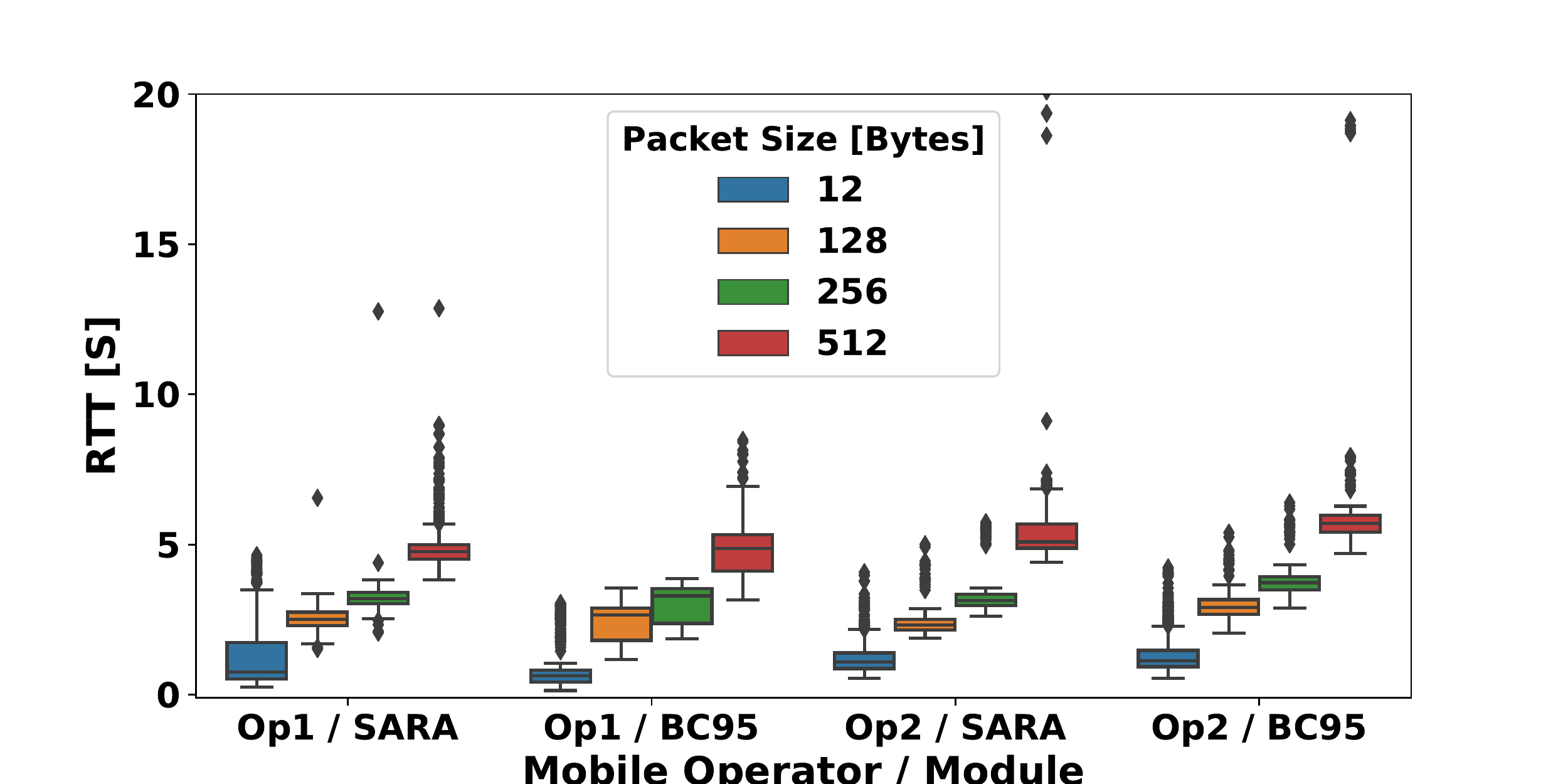}
    \caption{\gls{rtt} per packet size under good signal.}
    \label{fig:rttVpacketSize} 
    \vspace{-2em}
\end{figure}

\noindent{\bf \gls{rtt}.}
We measure \gls{rtt} through the device logs.
Fig.~\ref{fig:RTTVlocationCoverage} and~\ref{fig:rttVecl} present how signal
quality affects \gls{rtt}.
Under good conditions, despite the differences in energy consumption presented
in Section~\ref{sec:energy_consumption_connect},
round trip delays show small variability among the combinations examined.
On the other hand, higher ECL values increase the duration of both
transmission and reception, due to the big number of repetitions, consequently
increasing \gls{rtt}.
In Fig.~\ref{fig:RTTVlocationCoverage}, the higher values of delay under bad
signal conditions for \telia compared to \telenor are attributed to the much
more frequent use of ECL 2. Fig.~\ref{fig:rttVpacketSize} presents the effect
of packet size on \gls{rtt}, where we observe \telia having a bit longer delay
than \telenor.
As expected, longer packets require more transmission and
reception time, increasing \gls{rtt}.

\begin{figure}
	\includegraphics[width=1.1\linewidth]{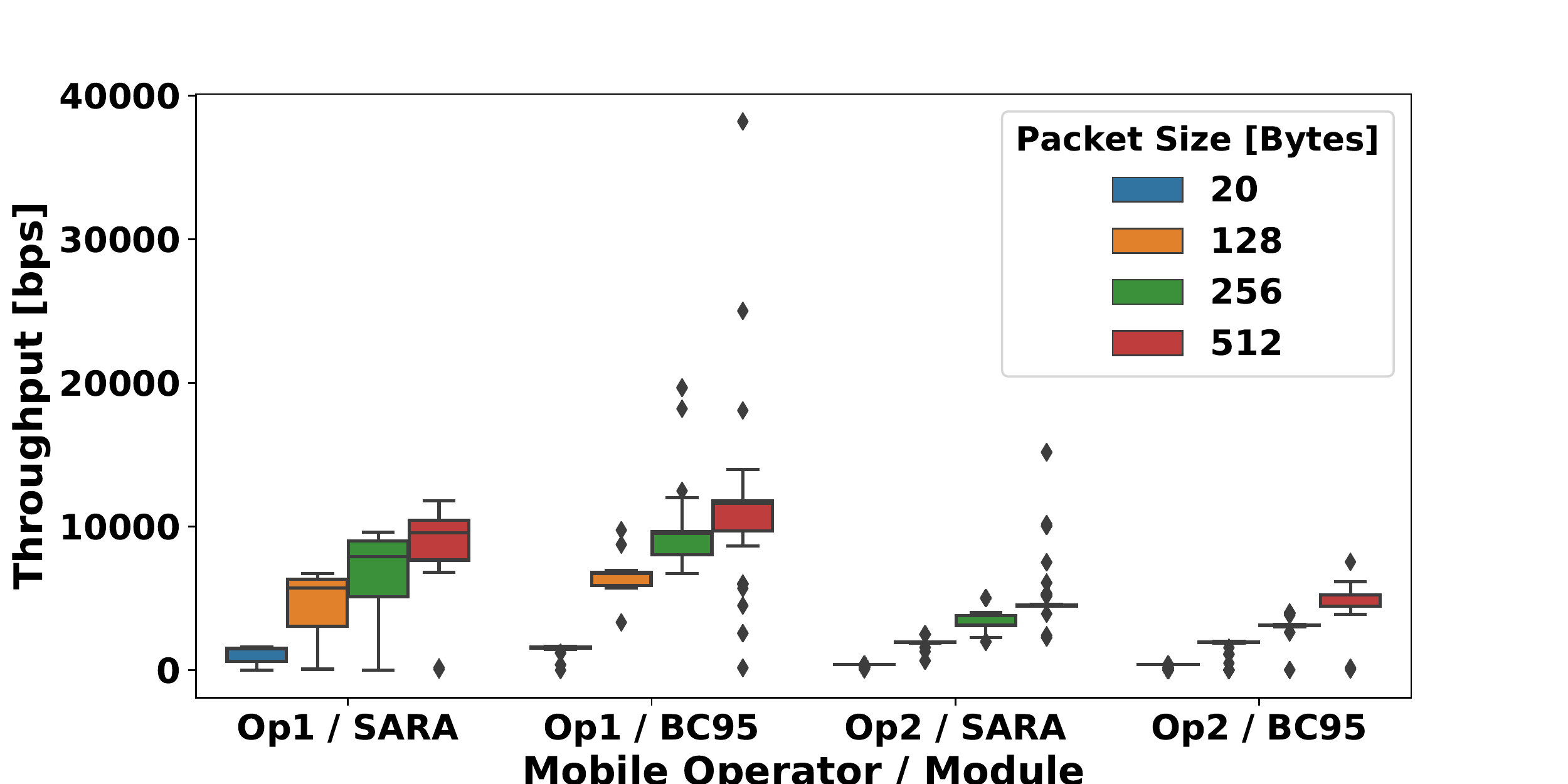}
	\caption{Throughput per packet size in good locations.}
    \label{fig:throughputVpacketSize}
    \vspace{-1.5em}
\end{figure}

\begin{figure}
	\includegraphics[width=1.1\linewidth]{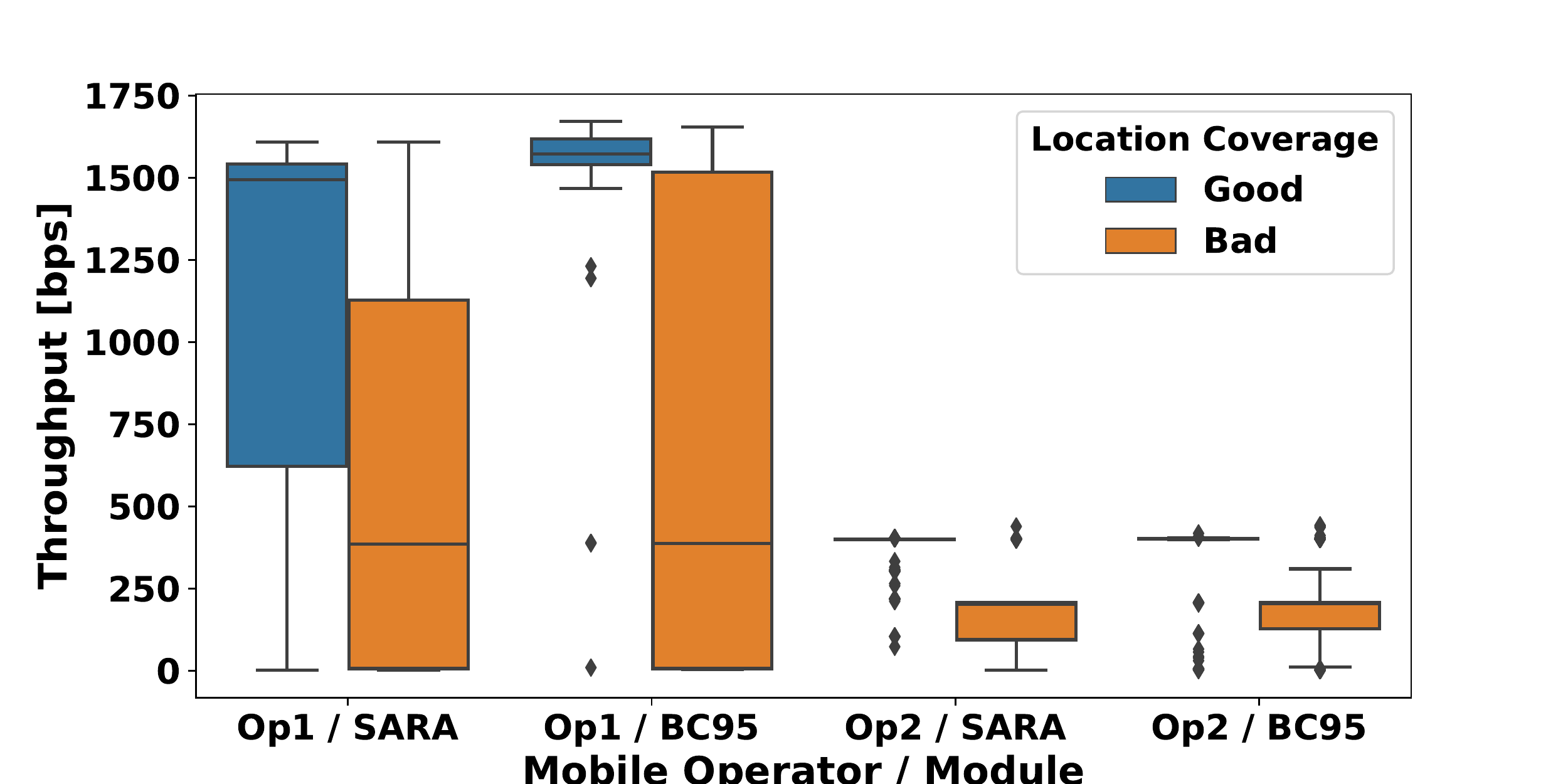}
	\caption{Comparison of Throughput for 20 bytes trasmissions, based on
		signal location.}
    \label{fig:throughputVlocationCoverage}
    \vspace{-1.5em}
\end{figure}

\noindent{\bf Throughput.}
Fig.~\ref{fig:throughputVpacketSize}
and~\ref{fig:throughputVlocationCoverage} break down the parameters that affect
throughput in the \gls{ul}. In our calculations, transmission starts from the
scheduling time and ends when the packet is transmitted. Due to the signaling
overhead, larger packets tend to be have a higher average transmission speed, as
shown in Fig.~\ref{fig:throughputVpacketSize}. Both operators are using 15~KHz
singletone mode, which has a
theoretical maximum \gls{ul} peak rate for Cat-NB1 devices of 16.9 Kbps. We
observe \telia being significantly slower than \telenor, even in good locations,
indicating inefficiencies in the signaling procedures and only \telenor
consistently gets measurements close to the theoretical maximum.
Signal quality has a great effect in measured speed, with experiments in
bad coverage locations resulting in less than half the speed.

\noindent{\bf Takeaways.} The NB-IoT networks we measure have higher packet loss
rates than ordinary LTE networks. ECL and packet size are the main factors
affecting \gls{rtt}, since they increase the time of all the RAN procedures.
Throughput is affected primarily by the operator and the packet size.

%% file: resultsSummary.tex
\section{Results Summary}
\label{sec:resultsSummary}

\begin{table}[h]
    \caption{Parameter importance hierarchy per state.}
        \label{tb:parameterImportance}
        \centering
        \footnotesize
        \begin{adjustbox}{max width=\textwidth}
        \begin{tabular}{ll}
            \toprule
            State                  & Parameter Importance  \\
            \midrule
            PSM                    & 1) Module                               \\
            \midrule
            \gls{edrx} - sleep     & 1) Module                               \\
            \midrule
            \gls{edrx} - listening & 1) Signal Quality 2) Module 3) Operator  \\
            \midrule
            Connected              & \pbox{5cm}{1) RAI flag 2) Signal Quality\\3) Module 4) Operator (assuming no bugs)\\\textcolor{red}{5) Packet Size (Only with RAI flag)}} \\          
            \bottomrule
            \end{tabular}
        \end{adjustbox}
    \end{table}

The previous Sections have presented our comprehensive measurement campaign,
over real commercial networks in a variety of conditions, devices and
configurations.
We have shown that NB-IoT offers greater configuration flexibility compared to
LTE, due to the recent NB-IoT specific enhancements and that
there are clear differences between operators and modules.
Most of the energy per transmission is consumed during the Active Timer phase,
thus it is critical for application developers to consider using the
appropriate RAI flag.

Operator misconfiguration may waste significant amounts of energy and it is
hard to detect without low level study of the energy traces.
The devices consume disproportionate energy, while meeting their expected
network KPIs (\ie RTT, packet loss), therefore there is no indication
of a problematic behavior.
For example, \telenor did not enforce quite periods during the Active Timer and
\telia consumed more power under \gls{rai} because of a bug that frequently
ignored the flag.

Both operators perform reasonably well under poor coverage.
The difference in median energy consumption between well and poorly
covered locations is attributed to ECL choice.
Operators should carefully select their ECL thresholds to avoid wasting energy.
Packet loss is very rare, even when operating in extremely challenging
conditions, where normal LTE devices would be ``out of signal''.
We could not find a correlation between packet loss and ECL level.
Instead, the operator with the more aggressive ECL thresholds exhibits higher
packet loss, while also suffering from increased RTT due to the high number
of repetitions associated with transmissions with \eclTwo.
Payload size only matters if \gls{rai} is in use and the dominant factor dictating
energy consumption is the module choice, especially under poor coverage.

Table~\ref{tb:parameterImportance} attempts to prioritize the impact of several
factors affecting energy consumption.
This table can be used as a starting point for building a power model estimating
an NB-IoT device's lifetime.
The choice of the module is a decisive factor in both Idle and Connected states.
The operator choice has the least, but still measurable, importance, unless
there are bugs present or the devices operate in locations with very poor
signal, where fine tuning of ECL thresholds is important.
Based on the above, applications developers should carefully select their
platform.

In brief, we conclude that NB-IoT deployments need
careful parameter tuning.
The use of Active Timer should be thought through.
A slight misconfiguration can result in excessive energy consumption and finally,
the measured NB-IoT deployments seem to fare well under poor coverage conditions
except for extreme cases.

%% file: discussion.tex
\section{Discussion}
\label{sec:discussion}

\noindent{\bf Device lifetime.}
In order to have an estimation of the device lifetime for a given battery
capacity, network configuration, and transmission frequency, we need to quantify
the energy consumption for the three distinct states of an \gls{nbiot} device
lifecycle: 1) \gls{psm}, 2) \gls{edrx} and 3) Connected state.
Thus, the expected lifetime $T_{lifetime}$ of the device, assuming no battery
degradation and a fixed transmission interval $T_{ti}$ is:

\begin{equation}
    \label{eq:lifetime}
    T_{lifetime} = \frac{E_{batteryCapacity}}{(E_{Con} + E_{eDRX} + E_{PSM})}*T_{ti},
\end{equation}

We sketch a toy example, to explore how different configurations and the choice
of the \gls{ue} impact device lifetime. In this example, a \gls{ue} under good
coverage sends a 20-byte UDP packet to an echo server, which responds to it.
This activity is repeated in 3 different intervals, with each interval being
representative of an NB-IoT usecase. The intervals are: i) 1h (\eg environment
monitoring), ii) 4h (\eg irrigation) and iii) 24h (\eg vehicle
automation)~\cite{mocnejnetwork}. The \gls{ue} spends the rest of the day in
Idle state. We explore two different configurations: default timers
(\ie the RAI flag is not set) and RAI-400.
We make two simplifications. First, we ignore the
energy consumed in the \gls{edrx} mode. This is a reasonable assumption for a
big number of \gls{nbiot} usecases, where a sensor reports data (Uplink), but is
not needed to be contacted (Downlink), thus \gls{edrx} can be disabled. Second,
we ignore the energy consumption associated with periodic \gls{tau} updates.
Given the frequency of Uplink messages, TAU is not needed in this scenario.
According to 3GPP's objectives for \gls{nbiot}, devices should be able to
achieve ``up to ten years battery life with battery capacity of 5 Wh
(Watt-hours), even in locations with adverse coverage
conditions''~\cite{3gpp45820}.
Thus, we assume a 5 Wh (18000 Joule) battery.
To estimate $E_{Con}$, we use the median values from
tables~\ref{tb:uplink_inactivity_good} and ~\ref{tb:uplink_rai_good}, which is a
good approximation given the compactness of the respective distributions.
The energy consumption during $PSM$ is calculated by multiplying the median
power consumption values from Sec.~\ref{sec:energy_consumption_idle_psm} with
the duration spent in Idle state.

\begin{table}[]
    \caption{Expected battery lifetime, in years, grouped by
        transmission intervals, for good signal conditions.
        The transmission is a single 20 byte UDP packet, which is echoed back.}
    \label{tb:battery}
    \centering
    \small
    \begin{tabular}{ll|ccc|ccc}
        \toprule
        Module                      & Op.      & \multicolumn{3}{c|}{Default timers}   & \multicolumn{3}{c}{RAI-400} \\
        \multicolumn{2}{c|}{}                  & 1h  & 4h  & 24h                       & 1h   & 4h   & 24h           \\
        \midrule
        \multirow{2}{*}{$\quectel$} & \telenor & 0.8 & 3.2 &  9.9                      &  6.1 & 25.5 & 45.4          \\
                                    & \telia   & 2.4 & 8.5 & 13.0                      &  6.4 & 30.1 & 47.6          \\
        \cline{1-8}
        \multirow{2}{*}{$\ublox$}   & \telenor & 0.5 & 1.9 & 6.0                       &  6.0 & 18.5 & 44.1          \\
                                    & \telia   & 1.6 & 5.9 & 5.7                       &  5.9 & 17.7 & 43.3          \\
        \bottomrule
        \end{tabular}
        
    \vspace{-1.5em}
\end{table}

Table~\ref{tb:battery} shows the expected battery lifetime in years for
different operator, module and configuration combinations. Misconfiguring energy
saving procedures, for example the lack of \gls{cdrx} in early measurements of
\telenor, drastically reduces the expected lifetime. Using RAI leads to
significant energy saving extending the battery lifetime by several years. Even
in this favorable scenario (\ie good signal, small packets), the use of RAI is
necessary to achieve the 10 year lifetime goal of 3GPP. Also, the differences
between modules translates to months of difference in battery lifetime, even in
the 1h interval scenario. Note that most of the energy consumption takes place
while the \gls{ue} is in deep sleep, because it spends the bulk of its lifetime
in that state. We have also evaluated other experiment conditions to gauge their
impact on battery lifetime. Taking the best case in Table~\ref{tb:battery}
above, that is $\quectel$ with \telia, we increase the payload size to 512
bytes, which consumes 0.20 J per message for RAI-400. The expected lifetime per
interval becomes: i) 8.6 ii) 23.3 and iii) 44.2 years. If we further assume bad
signal conditions, with RAI-400 and payload 512 bytes, the median consumption
becomes 3.09 J, thus making the expected lifetime be i) 0.7, ii) 2.5 and iii)
12.3 years.

The above logic may be applied to a plethora of other NB-IoT usecases.
For example, a home alarm is expected to be in Idle state, until
a very infrequent trigger (\eg once every three months), that is safe to ignore
in our calculations.
In this case, we would have to take into account the periodicity of TAU updates,
($T_{TAU}$)
which energy-wise can be assumed to be equal to a 20 byte packet transmission
with RAI-200, as these are the only other events that provoke a transition to the
Connected state.
The expected lifetime would be given by Equation~\ref{eq:lifetime} with 
$T_{ti} = T_{TAU}$.
On the other hand, a door lock use case would be dominated by random triggers.
We can get an estimate for the $T_{ti}$ from the expected number of triggers per
week.
Equation~\ref{eq:lifetime} is not affected by the distribution of $T_{ti}$, so
using the average value is sufficient.
In both use cases, we would also be interested in one-way delay and packet loss.
As we have shown, RTT is rarely above 10 seconds, even in extreme cases, and we
may ensure delivery through an application layer protocol.
In the rare Downlink-heavy use cases, such as unlocking city bikes, the dominant
factor in Equation~\ref{eq:lifetime} would be $E_{eDRX}$, because continuous
reachability is needed and thus, the heavy use of \gls{edrx} is a necessity.
\gls{edrx} is given by Equation~\ref{eq:edrx} and $T_{ti}$ could again be an
average value based on the expected number of triggers per week.

Based on the above, the use of default timers should be carefully thought
through, employing the \gls{rai} flag whenever possible. Any use case that does
not involve multiple communication from the server side, following the
initiation of an \gls{ul} transmission, should do away with it. It is the
default configuration, however, which means that most users might end up using
it unknowingly. It is not reasonable to assume that application developers will
be well versed in all aspects of energy saving in \gls{nbiot}. They, however,
need to familiarize themselves with the terms in Equation~\ref{eq:lifetime}.
Furthermore, the \gls{ue} vendors need to publish power ratings for their
devices when in deep sleep. Operators need to publish details on how they
implement energy saving and to certify common \gls{ue}s chipsets. The
availability of such information will make it easier for use case owners to come
up with reasonable battery lifetime estimates.

\noindent{\bf Feedback to operators.}
We have reported our findings to both operators, which they have fortunately
taken into account.
\telia had a bug with RAI-200, that was fixed after reporting it during our
main measurement campaign, achieving 80\% better energy efficiency
(see Sec.~\ref{sec:connected-rai}).
During the measurement period, \telenor did not support some \gls{nbiot} power
saving mechanisms, resulting in higher energy consumption than \telia.
During the Inactivity timer period, instead of performing \gls{cdrx}, the modules
were constantly on a high power paging state.
We have informed \telenor of this anomaly, they later informed us that it has
now been fixed.
We have then collected a complementary dataset in the first half of July 2019,
where we observe clear improvements.
The new measurements confirm that \telenor now implements \gls{cdrx}.
In these experiments we use SARA-N2 to send 20 bytes, from a location
with good coverage, using the default timers. 
The median energy consumption of the Connected state is now 0.912 Joule, having
improved by 77\%.
Actually, the energy consumption has become lower than \telia's, because the
\gls{cdrx} mode of \telenor, has fewer and more spaced out listening occasions.
\telia supported these power saving features from the beginning of our
measurements, thus we do not observe any differences at the newer dataset.
The immediate impact of our study, highlights the need for similar studies as
\gls{nbiot} is being rolled out and soon 5G will be.

\noindent{\bf Possible implementations of this study.}
To the best of authors' knowledge, literature research is more focused on optimizing energy efficiency through
improved scheduling strategies or predicting traffic needs, which have general applicability
\cite{lee2017prediction, liu2018energy, oh2016efficient}.
Instead, this work is based on empirical evaluation of real deployments, which shed light
on the relation between energy consumption and latency and UE/network settings. Further studies and investigations can use this
information as a guideline to design algorithms or mathematical models to define the 
optimal parameter configuration with the correct  metadata
metrics (\ie ECL, SNR, RSRP) as input.
These algorithms could be applied by networks operators to tune the UE/network
settings to fit a specific type of traffic, transmission frequency
and requirements in an \gls{iot} application, increasing both user experience and
network efficiency.

\noindent{\bf Earlier studies.}
The closest works to ours
are~\cite{martinez2019exploring, yeoh2018experimental, duhovnikov2019power},
which use the same devices but over different networks and in a smaller variety
of scenarios.
\cite{martinez2019exploring} performed measurements over a single commercial
network in Spain with the same devices.
Similarly to us, they observe that Quectel has better energy consumption to
Ublox and that packet size does not affect energy consumption.
In contrast, they report significant gains by using the RAI flag only for ublox,
and considerably less energy needed to listen for PO during \gls{edrx} for both
modules.
We expand on their work, by comparing the
performance of two operators and attempt to identify the parameters mostly
affecting lifetime.
In~\cite{yeoh2018experimental}, they use the same devices, but with older
firmware that supported only release 13 features.
The experiments reported are an integration study for the network of Telekom
Malaysia, where they also study energy efficiency.
They reach the same conclusion as us, that in order to achieve the promised
lifetime a careful set up of the NB-IoT device's firmware is necessary.
In contrast, we perform our experiments over commercially available networks
with the latest firmware of both devices that supports all the
current-generation power saving features and under a variety of signal
conditions.
In~\cite{duhovnikov2019power}, authors perform a small scale experiment to
measure the expected lifetime of an NB-IoT device, based on SARA-N2, in the
context of aviation use cases.
Their testbed connects over a private and two commercial networks and they
discover that using PSM in Idle state has the highest impact on achievable
battery lifetime. 
Compared to the 3 above studies, our experiments are more thorough and use
the latest NB-IoT features commercially available.
We further attempt to provide explanation of the artifacts we observe, identify
key parameters for enhancing lifetime and cooperate with operators to improve
their networks.

%% file: related_work.tex
\section{Related Work}
\label{sec:related_work}

In addition to the early power measurement studies presented in
Section~\ref{sec:discussion}, various works have attempted to model NB-IoT power
consumption and device lifetime. The authors of \cite{andres2019analytical,
  andres2017optimized} presented a Markov chain analysis of the average energy
consumed to transfer one uplink report using the Control Plane procedure.
In~\cite{aalborgMasterThesis} an emulator has been used to create an empirical
lifetime model, based on device configuration. The same testbed has been used
in~\cite{lauridsen2018empirical}, where authors measured two early NB-IoT device prototypes and used the results to make lifetime estimation
projections. An early simulation study of various IoT technologies' coverage,
including NB-IoT, based on a Danish region's topology has been presented
in~\cite{vejlgaard2017coverage}. In~\cite{sultania2018energy} Sultania et al.
proposed an analytical model to estimate the average energy consumption of an
NB-IoT device using the Release 14 power saving enhancements. The work
in~\cite{harwahyu2019repetitions} presented an analytical model to explore the
trade-offs between repetitions and the built-in MAC layer retransmission
mechanism of LTE, concluding that fewer repetitions with more retransmissions
achieves higher successful probability. Recently,~\cite{feltrin2018nb} presented a
theoretical mathematical model to predict performance and propose optimal
network configuration in different scenarios.
El Soussi et al.~\cite{el2018evaluating} evaluated the overall performance of
NB-IoT in the context of a smart city. They proposed a theoretical model for
calculating the energy consumption and conclude that a lifetime of 8 years is
possible, under poor coverage, while sending one message per day. Finally, the
authors of~\cite{mozny2019performance} studied the relationship between signal
strength and delay through a small number of experiments performed in a
laboratory testbed measuring a device prototype based on the SARA-N2 module.
These works rely either on emulating parts of the network or simulations. In
contrast, we perform large scale experiments in the wild using two different modules and two operators, and compared to~\cite{martinez2019exploring, yeoh2018experimental, duhovnikov2019power}, discussed in the previous section, our work is more thorough and uses the last NB-IoT features available. Finally, to the best of our knowledge, there is no other empirical study on NB-IoT packet loss under real conditions.

%% file: conclusions.tex
\section{Conclusions}
\label{sec:conclusions}

We conducted a comprehensive measurement study of the energy consumption of
two popular \gls{nbiot} boards that connect to two commercial deployments in a
European country.
Our findings indicate that \gls{nbiot} is far from being plug and play and
requires careful setting for improving energy efficiency.
Since we focus on configuration parameters and their impact on the energy consumption,
our recommendations can be generalized to any NB-IoT deployment.
We observe that the main factors determining energy consumption and thus
battery life are: 1) module, 2) operator, 3) signal quality, 4) use of energy
saving enhancements such as RAI and eDRX and, 5) in a limited number of
scenarios, packet size.
Furthermore, our analysis has helped the measured networks identifying and
fixing a couple anomalous configurations, and we could finally track the
effectiveness of this adjustments.
Finally, we have indicated strategies for improving energy efficiency, pointing
out the elements that could bring to energy waste without improving the reliability, 
such as too aggressive \gls{ecl} thresholds or not using the \gls{rai} flag.
We also identified the key parameters needed for estimating the battery lifetime, 
and which of the metadata reported by the device are more meaningful.
Possible future research directions include the energy impact of application
layer protocols such as MQTT and CoAP, as well as recommendations for parameter
tuning of these protocols.





%% file: ack.tex
\section{Acknowledgments}
\label{sec:ack}

This work has been supported by the European Community through the 5G-VINNI
project (grant no. 815279) within the H2020-ICT-17-2017 research and innovation
program.

%% file: dataPreprocessingAndMetadata.tex
\section{Data pre-processing}
\label{sec:dataPreprocessingAndMetadata}

\begin{figure*}
    \centering
    \begin{subfigure}[t]{0.49\linewidth}
    \centering
    \includegraphics[width=\linewidth]{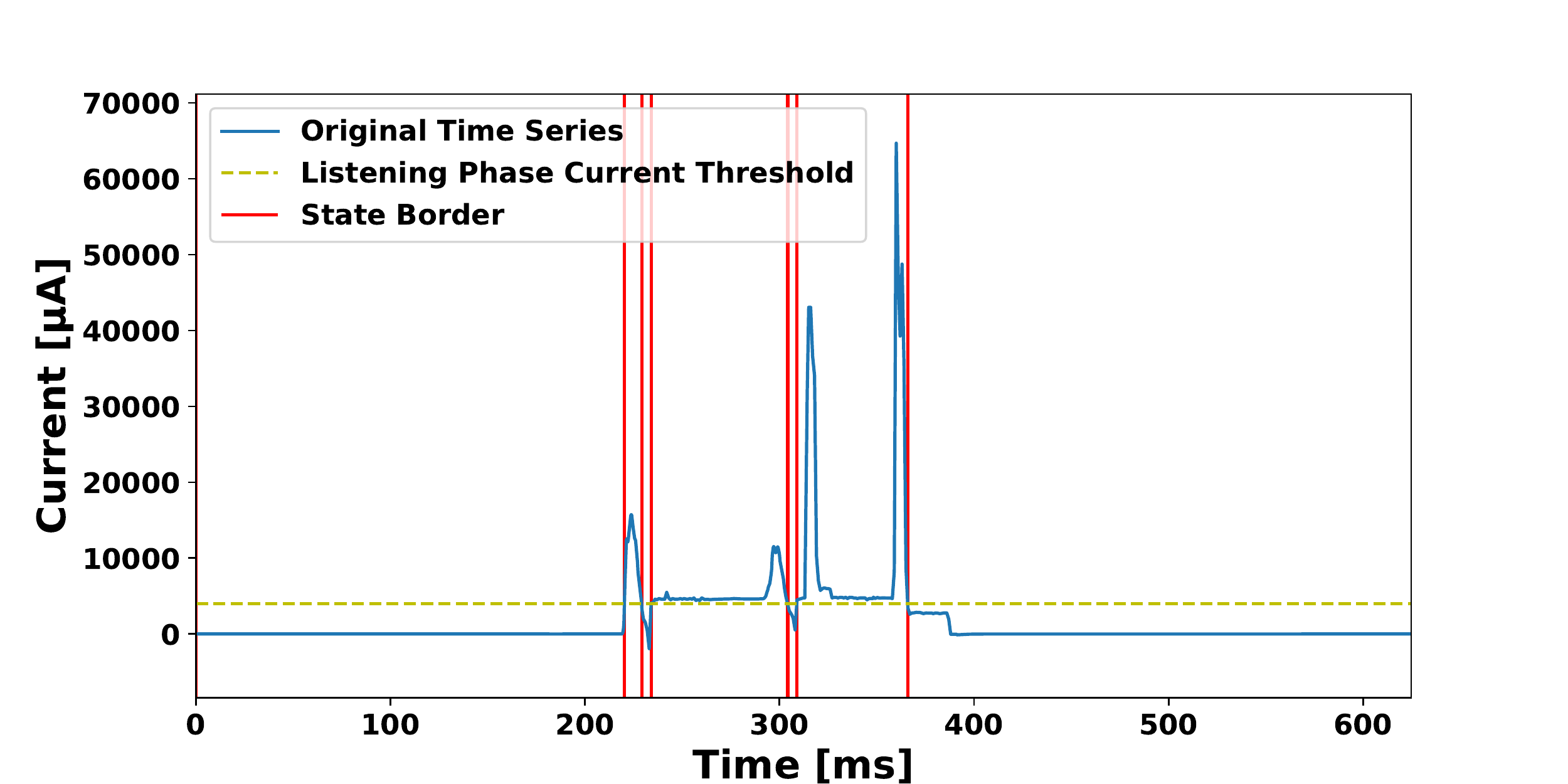}
    \caption{Detection based on the original time series: $T$.}
    \label{fig:dataPreprocessing_naive}
    \end{subfigure}
    \begin{subfigure}[t]{0.49\linewidth}
    \centering
    \includegraphics[width=\linewidth]{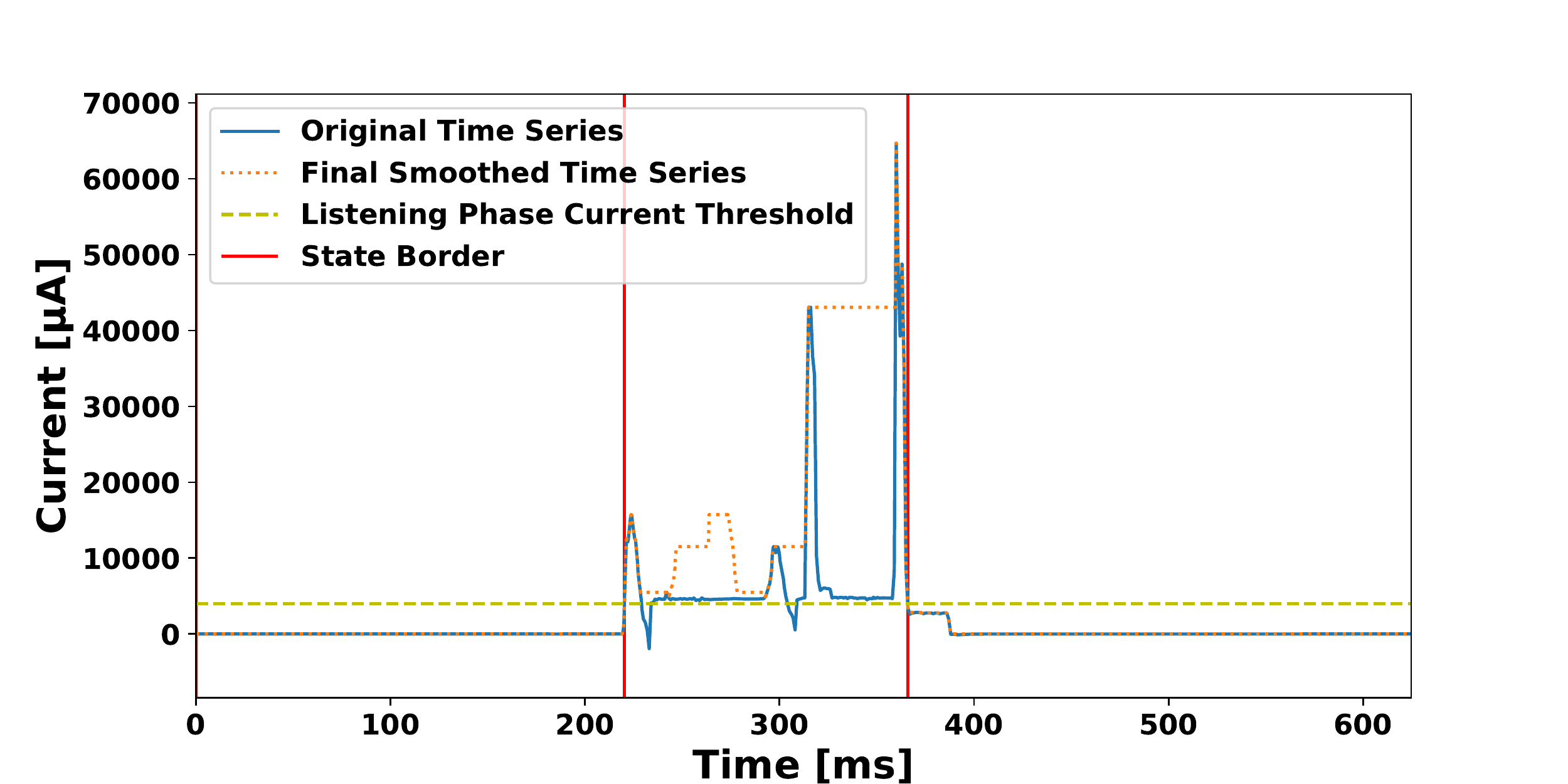}
    \caption{Detection based on the final smoothed time series: $FSTS$.}
    \label{fig:dataPreprocessing_smart}
    \end{subfigure}
    \caption{Results of the phase detection algorithms when trying to detect an
    \gls{edrx}  listening phase.
    The red vertical lines signify the borders between different phases.}
    \label{fig:dataPreprocessing}
    \vspace{-1.5em}
\end{figure*}

In this Appendix we present how we synchronize the logs of the Otii Arc power
measurement device with the logs of the \gls{ue}s.
The \gls{ue}s report network metadata such as \gls{rrc} connection state and DRX.
These must be
synchronized with the Otii power measurements to avoid misattributing energy
consumption, connection state-wise. A listening phase in Idle state typically
lasts less than 300 ms and a \gls{cdrx} one less than 30 ms, thus the
synchronization ideally should have an error of a few ms at most.
Unfortunately,
the \gls{ue} and the power meter clocks could not be synchronized to the
required accuracy. Instead, we resort to time series analysis to dissect the
current consumption time series into phases.
We leverage the fact that the power
consumed in different phases is markedly different, as well as characterized by
different patterns (see Sec.~\ref{sec:nbiot_primer}).
For example, we are able to isolate
the DRX listening phase or the synchronization procedure.

Fig.~\ref{fig:dataPreprocessing} presents an example of how our phase
detection algorithms operate when detecting an \gls{edrx} listening phase. Other
events are detected in a similar fashion. Depending on which phase we are trying
to detect, we can set a current threshold, above which we assume the device is
within that phase. Thus, the edges of each phase are the points where the time
series cross this threshold. The threshold is determined by the value of the
95th percentile of the current of a typical phase. As can be seen in
Fig.~\ref{fig:dataPreprocessing_naive}, the original power monitor time series
($T=\{T_{1},\dots,T_{n}\}$, where $n$ is the number of observations) is very
volatile, crossing the threshold multiple times within a single phase, which 
makes phase detection hard. Based on $T$, we create two smoothed time series,
each aimed to properly detect one edge of the target phase. The window size of
the smoothing functions is determined empirically per measurement at a value
that removes fluctuations, while avoiding overlap with neighboring listening
phases. These are combined to create a \texttt{Final Smoothed Time Series}: $FSTS$,
where both edges of every phase are well defined and we use this as a guide to time
stamp the start and end of each phase. Then, we use the original time series $T$
to get the energy consumption and the rest of the target metrics of the now well
defined phases.

For example, to identify the \gls{edrx} listening phase, of
Fig.~\ref{fig:dataPreprocessing} we have to calculate: A) The moving max of
all the values ahead of the current one in the window, aimed to detect the phase
end. Each element of this time series: \texttt{Moving Max Forward}: $MMF$, is given by: $MMF_{i} =
max\{T_{i},\dots,T_{i+W}\}$, where $W$ is the window size of the function. B)
The moving max of all the values before the current one in the window \texttt{Moving Max Backward}: $MMB$,
aimed to detect the phase start. The elements of $MMB$ are given by: $MMB_{i} =
max\{T_{i-W},\dots,T_{i}\}$.
C) Calculate the $FSTS$, by overlapping $MMB_{i}$ and $MMF_{i}$.
We do so by taking their minimum: $FSTS_{i} = min(MMB_{i},MMF_{i})$. 
$FSTS$ has the
property of increasing as soon as the current increases, while not being
sensitive to current fluctuations, thus creating a tight mask around the phase
we want to detect.
The resulting $FSTS$ and the phase borders it generates for
the \gls{edrx} listening phase detection example are seen in
Fig.~\ref{fig:dataPreprocessing_smart}. Finally, we apply some filtering on
the detected events, to remove artifacts, such as current spikes when we poll
the modules for metadata.

Detection of other events, with more distinct patterns, such as the transition
between Connected and Idle state, can be simpler.
For example, to identify
Connected and Idle states, a single smoothed time series of a moving median
around the central value of the window is enough to properly identify both the
beginning and the ending of a state. 
This is possible due to the bigger difference in the power
consumption between the two states and the bigger duration and periodicity of
these events.

The smoothing functions used depend on the event. The parameters depend on the
behaviour of the current time series, which is affected by experiment conditions
and settings, thus might need adjustment per measurement. An added benefit of
this method is that it is very computationally efficient, since it utilizes time
series libraries instead of loops, providing fast results in processing the very
big files provided by the power monitor tool.

%% file: snr_rsrp_mapping.tex
\section{SNR to RSRP mapping}
\label{sec:snr_rsrp_mapping}

NB-IoT devices calculate \gls{sinr} over the whole 180 KHz
channel: $SINR=\frac{12*RSRP}{I_{tot}+N_{tot}}$.
In contrast, \gls{rsrp} is calculated over a a single Resource Element (RE),
which has 15KHz bandwidth and is assumed to be free of noise and interference.
Thus, to map \gls{sinr} and \gls{rsrp} we need to modify the above equation to
take into account only 15KHz: $SINR_{15KHz}=\frac{RSRP}{I_{15KHz}+N_{15KHz}}$.
In our experiments, due to the limited adoption of NB-IoT and the nature of
GuardBand deployment, we can safely assume that interference is minimal,
especially in the poor coverage scenarios, thus:
$SNR_{15KHz}=\frac{RSRP}{N_{15KHz}}$
The $N_{15KHz}$ depends on the thermal noise density and the receiver noise figure,
which have typical values of $N_{thermal}=-1740 cBm/Hz$ and
$NF_{receiver}=70cB$, respectively. Thus the thermal component of the noise is:

\begin{equation}
\begin{aligned}
\label{eq:thermalNoise}
N_{thermal\_15KHz}=-1740cBm/Hz+100log(15000Hz) \\
\approx-1322cBm.
\end{aligned}
\end{equation}

$N_{15KHz}$ then becomes:

\begin{equation}
\begin{aligned}
\label{eq:noiseRE}
N_{15KHz} = N_{thermal\_15KHz} + NF_{receiver} \\
= -1322.39 cBm + 70 cB = -1252 cBm.\\
\end{aligned}
\end{equation}

Finally the ideal mapping of \gls{snr} values to \gls{rsrp} under our
assumptions in logarithmic scale is:

\begin{equation}
\begin{aligned}
\label{eq:rsrpsnrmap}
SNR_{15KHz}=\frac{RSRP}{N_{15KHz}}\Rightarrow \\
SNR_{cB}=RSRP_{cBm}+1252.
\end{aligned}
\end{equation}